\documentclass[twocolumn, aps, longbibliography]{revtex4-1}
\usepackage[utf8]{inputenc}
\usepackage{amsmath}
\usepackage{graphicx}
\usepackage{subfiles}
\usepackage{color}

\usepackage{ulem}

\makeatletter
\begin{document}
\title{Predicting phase preferences of two-dimensional transition metal dichalcogenides using machine learning}


\author{Pankaj Kumar$^{1}$}
\author{Vinit Sharma$^{2}$}
\author{Sharmila N. Shirodkar$^{1}$}
\author{Pratibha Dev$^{1}$}
\email{pratibha.dev@howard.edu}

\affiliation{$^{1}$Department of Physics and Astronomy, Howard University, Washington, D.C. 20059, USA.\\
$^{2}$National Institute for Computational Sciences, Oak Ridge National Laboratory, Oak Ridge, TN 37831, USA}
\begin{abstract}


Two-dimensional transition metal dichalcogenides (TMDs) can adopt one of several possible structures, with the most common being the trigonal prismatic and octahedral symmetry phases. Since the structure determines the electronic properties, being able to predict phase-preferences of TMDs from just the knowledge of the constituent atoms is highly desired, but has remained a long-standing problem. In this study, we applied high-throughput quantum mechanical computations with machine learning algorithms to solve this old problem.  Our analysis provides insights into determining physiochemical factors that dictate the phase-preference of a TMD, identifying and going beyond the attributes considered by earlier researchers in predicting crystal structures. A knowledge of these underlying physiochemical factors not only helps us to rationalize, but also to accurately predict structural preferences. We show that machine learning algorithms are powerful tools that can be used not only to find new materials with targeted properties, but also to find connections between elemental attributes and the target property/properties that were not previously obvious. 

\end{abstract}

\date{\today}
\maketitle

\section{Introduction}


Layered transition metal dichalcogenides (TMDs), given by the formula MX$_2$ (M = transition metal; X = chalcogen), represent a chemically well-defined family of materials. 
As bulk crystals, TMDs are formed from weakly-bonded van der Waals stacks of MX$_2$ layers, and were studied extensively in the 1960s --1980s~\cite{Wilson_Yoffe_1969,Yoffe_review1973,Gamble1974,Hoffmann_1984}. Interest in TMDs was revived nearly two decades ago when it was realized that bulk TMD crystals can be separated into two-dimensional (2D) monolayers and few-layer thick crystals~\cite{Novoselov2005}.  
Hence, a considerable experimental effort has been dedicated to their synthesis~\cite{Ayari_2007_JAP,Matte_2010,Coleman2011,Liu_MoS2_2012,Nakano_WSe2_2017,Kaur2018,Puthirath_JPCC_2021}. 
The diverse electronic, magnetic and topological properties of TMDs~\cite{Chhowalla2013,Qian2014,Xiao2012,Mak2012,Ravindra2019} makes them attractive for a number of technologies~\cite{Wang2012,Lin2014,Hayat2017,Duerloo2014,Li2016,Yang2017,Rhodes2017,Wang2017,Loncar_TMD_Nature_2017,Manchanda2020,Manchanda_PtSe2_2021}. 
With the recent creation of novel Janus TMD crystals, MXY, in 2017 (X and Y are two different chalcogens on two faces)~\cite{Lu2017}, it has become clear that the TMD family is growing larger, along with the list of their properties, which now includes ferroelasticity and ferroelectricity~\cite{Lu2017,MoSSe_Zhang_2017,Zhang_VSSe_2019}.  


Unlike 2D layered materials such as graphene and hexagonal boron nitride, TMDs can exist in multiple crystal phases. Both traditional (MX$_2$) and  Janus (MXY) TMDs most commonly adopt either a trigonal prismatic phase (\textit{1H} phase) or an octahedral symmetry phase (\textit{1T} phase).  As the structure determines their properties, knowledge of factors that determine phase-preferences of TMDs is very important and has long been debated in the literature~\cite{Gamble1974, Hoffmann_1984}. These earlier works, using chemical and physical considerations, showed how the structural preferences are related to different factors, such as: (i)  the Pauling's cationic to anionic radius ratio~\cite{Pauling1929}, with the ratio obtained by combining the structural constraints with the the fractional ionicity of bonds within a TMD layer~\cite{Gamble1974}, and (ii) the d-electron counts of the transition metals~\cite{Hoffmann_1984}.  
All of these earlier studies involved traditional TMDs that were known at the time of the respective studies.  
Due to the rapid materials discoveries in recent years, which include discoveries of novel TMDs, it is important to develop phase-prediction methods that can predict preferred structures for a much larger chemical phase-space and are generalizable to even different classes of materials, while still having roots in physical and chemical intuition.



\begin{figure*}
    \centering
    \includegraphics[width=0.75\linewidth]{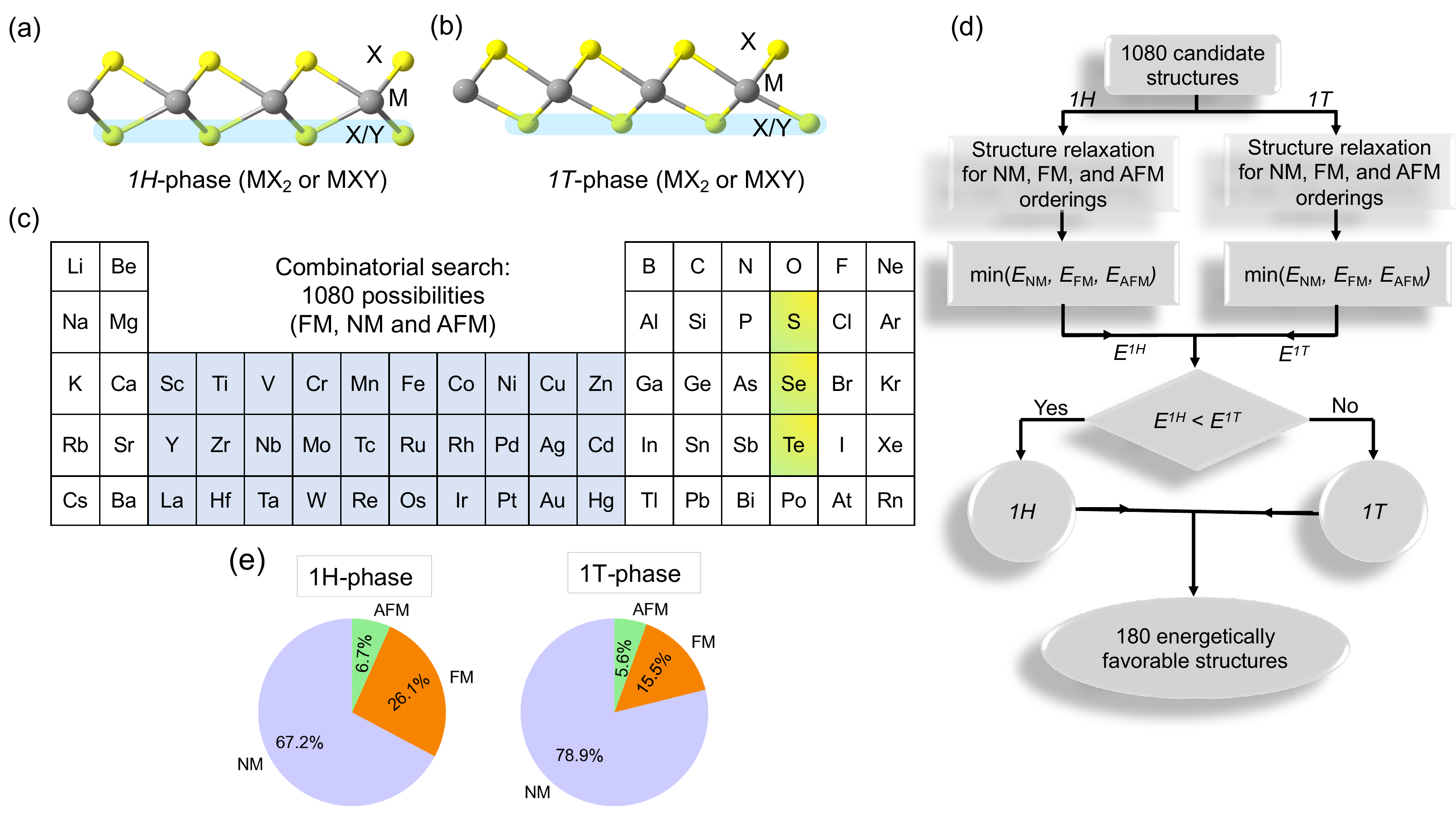}
    \caption{Schematic representation of (a) \textit{1H} and (b) \textit{1T} phase traditional and Janus TMD monolayers. The Janus structure consists of two different chalcogens (X and Y) on two faces of a TMD. (c) Periodic table highlighting the transition elements and the chalcogen atoms considered for the combinatorial study. (d) Flow chart of systematic combinatorial search of traditional and Janus monolayers. (e) Pie charts showing the \% distributions of the most stable magnetic orderings -- ferromagnetic (FM), antiferromagnetic (AFM), and non-magnetic (NM) -- for \textit{1H} phase and \textit{1T} phase TMDs.}
    \label{fig:schematics}
\end{figure*}

In this work, we created a generalizable framework for predicting phase preferences of TMDs by combining high-throughput quantum mechanical computations with machine learning (ML) algorithms (see detailed discussion in Supplementary Note I \cite{supp}). This powerful combination can help to explore the materials phase space more economically, and has been successfully deployed in several studies~\cite{Huan2016,Tawfik2019,Sharma2020,Masubuchi2020,Schleder2020,Choudhary2020,Jarvis2020}. We also went beyond the known TMDs, creating novel combinations of transition elements and chalcogen atoms, thereby expanding the chemical phase space over which our analysis is valid. 
Density functional theory (DFT) calculations were used to generate a database of formation energies ($\Delta  E_f$) for different TMDs. Machine learning models were then trained and tested for their ability to predict chemical stability and phase preference of the TMDs. 
We quantitatively show that the most important factors determining the formation energy, as well as the crystal structure of a TMD, are the relative elemental dipole polarizabilities, electronegativities, ionization energies and commonly exhibited valencies of the elements comprising the TMD structure. Two additional attributes that play a role (albeit to a lesser extent) are the electron-affinities and heat of gas-phase formation. We also found that swapping the dipole polarizabilities with covalent radii, and vice-versa, in our analysis yields nearly-equivalent ML models. This is not surprising because an atom's dipole polarizability is a size-dependent attribute and is a measure of the covalency of the bonds that it forms. The relative radii, degree of ionicity (as given by the difference of electronegativities) and relative valencies (related to d-electron counts), are expected to contribute to the relative stability of the \textit{1H} and \textit{1T} phases. 
These dependencies are in accordance with the works by Gamble~\cite{Gamble1974} and by Kertesz \textit{et al.}~\cite{Hoffmann_1984} who used chemical and physical arguments to rationalize phase preferences, while the other features represent previously unknown contributors.  Hence, by employing ML algorithms to find new connections between relative elemental attributes of the constituents and the target property/properties of the resulting compounds, we are able to both rationalize and predict chemical stabilities and structural preferences of the TMDs.



\section{Technical Details}

\subsection{High-throughput first principles calculations}

In this work, we used ML algorithms to identify relationships between phase-stability of TMDs and their respective chemistries in a high-dimensional space of independent variables. Since the accuracy of datasets as well as their sizes dictate the accuracy of ML models trained on these datasets, it was important to go beyond known TMDs.
Hence, we performed high throughput first principles calculations to generate both conventional and Janus TMDs [Fig.~\ref{fig:schematics}(a) and (b)] from different combinatorial possibilities of chalcogens (X,Y = S, Se, Te) and transition metal (TM) atoms belonging to the 3d-, 4d- and 5d-series [highlighted in Fig.~\ref{fig:schematics}(c)]. There were a total of 360 chemically- and structurally-distinct crystals (i.e \textit{1H} phase and \textit{1T} phase structures) formed from 30 transition elements (Sc-Zn, Y-Cd, and La-Hg), and 3 chalcogen atoms.  Since these structures can be magnetic, each TMD was further allowed to adopt ferromagnetic (FM), antiferromagnetic (AFM) and non-magnetic (NM) configurations, yielding a total of 1080 candidate structures [see Fig.~\ref{fig:schematics}(d)].  Our spin-resolved DFT calculations were performed using the Vienna \textit{Ab-initio} simulations package (VASP)~\cite{Kresse1994,Kresse1996}. The Perdew, Burke, and Ernzerhof (PBE) generalized gradient approximation~\cite{Perdew1996} was used to account for the exchange-correlation effects. In order to minimize interactions between the periodic images of the monolayers, a large vacuum of 20\,\AA{} was used in the direction normal to the 2D layers.  In order to determine if the TMDs will be magnetic, we calculated  the energy differences (per formula unit) between the non-magnetic and magnetic structures, where the latter can have either FM or AFM alignments of the moments. For the FM alignment, we performed a spin resolved calculation, wherein the moments on the atoms within a unitcell were initialized to non-zero values. This generated a ferromagnetic structure since the periodic boundary conditions ensured that the moments (if non-zero) in the cell and its images were all aligned.  To create magnetic structures with AFM alignment of moments, we doubled the size of the unit cell in one of the lateral directions, and initialized the moments on the magnetic species in the neighboring cells so as to align them antiferromagnetically~\cite{Dev_PRL_2008, Dev_PRB_2010}. All structures were optimized using an energy cutoff of 550\,eV and a $\Gamma$-centered \textit{k}-point mesh~\cite{Monkhorst1976}, which was equivalent to a 21$\times$21$\times$1-grid for a unit cell, until the forces on each of the atoms was less than 1\,meV/\AA{} and the total energy values converge to better than 10$^{-7}$\,eV.


Formation energies for different TMD monolayers were calculated using:
\begin{equation}
  \Delta  E_f = E_{MXY} - (E_M + E_X + E_Y)
    \label{eq1}
\end{equation}
Here, $E_{MXY}$ is the total energy of the monolayer. $E_M$, $E_X$ and $E_Y$ are the total energies for the the transition metals (M) and the chalcogens (X, Y), respectively, in their most stable bulk form. Formation energy is an indicator of a material's chemical stability, although it does not account for possible mechanical instabilities. For the latter, one needs to determine dynamical stability by calculating the phonon spectra, and thermal stability using \textit{ab-initio} molecular dynamics at different temperatures. It is also worth mentioning that some of the materials that may display mechanical instabilities as freestanding monolayers can possibly be stabilized on appropriate substrates. However, determining mechanical stability and/or conditions under which a monolayer can be stabilized is beyond the scope of present work.

\subsection{ML-based materials property prediction}

\subsubsection{Primary descriptor/feature dataset}

Choosing an optimal set of independent descriptors/features is of utmost importance in building effective machine learning models. 
As a first step, our goal was to find easily accessible descriptors that can uniquely describe the system and provide a meaningful mapping to the target property. To achieve this goal, we initially chose thirteen elemental properties of the atoms that form the TMD monolayers.  More specifically, the primary atomic and bulk elemental descriptors considered here are:  heat of gas-phase formation (HF), metallic radius (MR), specific heat (SP), covalent radius (CR), electron affinity (EA), atomic weight (AW), electronegativity (EN), heat of fusion (HS), atomic radius (AR), dipole polarizability (DP), ionization energy (IE), valency (VS) and Shanon ionic radius (IR) of the elemental species forming the TMDs~\cite{Shannon1976,mendeleev2014}. Since we studied compounds in this work, we actually needed compound descriptors/features that could describe a TMD with a particular combination of elements. Hence, instead of the aforementioned elemental attributes in their unchanged forms, we considered different combinations of the elemental descriptors to form compound descriptors.  We also refrained from using properties of TMDs themselves (e.g., lattice constants) that one obtains only after performing a DFT-calculation or an experiment on a TMD, which would have defeated the purpose of using machine learning. The compound descriptors formed from the elemental descriptors were divided into three sets: (i) Set A, formed by finding the ratio between the elemental attributes of the metal and the averaged attributes of the chalcogens forming each of the TMDs, such that a compound attribute in Set A is given by:  $\mathcal{F}_{M/\bar{X}} =2\,\mathcal{F}_{\textrm M}/(\mathcal{F}_{\textrm X}+\mathcal{F}_{\textrm Y}) = \mathcal{F}_{\textrm M}/\mathcal{F}_{\mathrm{\bar{X}}}$, where $\mathcal{F}_{\textrm M}$, $\mathcal{F}_{\textrm X}$, $\mathcal{F}_{\textrm Y}$ and $\mathcal{F}_{\mathrm{\bar{X}}}$ are the elemental attributes of atoms M, X, Y, and the averaged attributes of the chalcogen atoms [$\mathcal{F}_{\mathrm{\bar{X}}}=(\mathcal{F}_{\textrm X}+\mathcal{F}_{\textrm Y})/2$], respectively; (ii) Set B, obtained by finding the difference in the averaged attributes of the chalcogen atoms and the elemental attributes of M atoms, such that a compound attribute in this set is represented by $\mathcal{F}_{\mathrm{\bar{X}M}} =\mathcal{F}_{\mathrm{\bar{X}}}-\mathcal{F}_{\textrm M}$; (iii) Set C, obtained by finding the difference of the elemental properties of the two chalcogens involved in the formation of different TMDs, such that a compound attribute is given by $\mathcal{F}_{XY}=\mathcal{F}_{\textrm X} - \mathcal{F}_{\textrm Y}$. Sets A--C thus contained a total of 39 compound descriptors.
 In order to identify independent compound descriptors, we calculated the Pearson correlation coefficients between the features. Formally, the Pearson correlation coefficient (or bivariate correlation) between two variables is obtained by finding the covariance of the two variables and dividing it by the product of their standard deviations. The Pearson correlation coefficient, $\mathcal{P}$, when applied to the features is effectively obtained by taking the inner product:

\begin{equation}\label{eq:Pearson}
\mathcal{P} (\vec{u}, \vec{v}) = \frac{\sum_{j=1}^{N} [u_{j} - \bar{u}][v_{j} - \bar{v}]}{\sqrt{\sum_{j=1}^{N} [u_{j} - \bar{u}]^2}  \sqrt{\sum_{j=1}^{N} [v_{j} - \bar{v}]^2} }
\end{equation}
Here, $\vec{u}$ and  $\vec{v}$ are any two of the M descriptors, and are $N\times1$-dimensional vectors, $N$ being the total number of materials. The mean value of the descriptor $\vec{u}$ is given by $\bar{u}=(1/N)\sum_{j=1}^{N} u_{j}$, where $u_{j}$ is the $j^{th}$ component of $\vec{u}$.  As the inner product in Eq.~\ref{eq:Pearson} is normalized, the lower and upper bounds of $\mathcal{P}$ are $-1$ and $+1$, corresponding to strong negative and positive linear relationships between the features, respectively. A value of $\mathcal{P} = 0$ indicates that the features are completely independent. 


\subsubsection{Machine Learning Algorithms}


The down-selected features were used as inputs in supervised learning of formation energies. For this, we employed two ML algorithms -- Random forest (RF) and Kernel ridge regression (KRR).

\noindent \paragraph{Random forest algorithm:}
The RF algorithm was used for two tasks: (i) regression to predict formation energies of the TMDs, and (ii) classification to predict the phase of the TMDs~\cite{Breiman2001}. RF consists of constructing a large number of individual uncorrelated decision trees (models) that work as an ensemble~\cite{ensemble}. In order to ensure that the different models (decision trees) are uncorrelated, each independently-constructed decision tree randomly samples the dataset with replacement, 
thereby avoiding overfitting. 
The final prediction is made by aggregating, either by averaging or with a simple majority vote. This ML method has only two parameters: (i) the number of variables in the random subset at each node, and (ii) the number of trees in the forest.  Although RF is not very sensitive to the specific values chosen for these two parameters, it is generally desirable to optimize these parameters during the model training stages.

 \noindent \paragraph{Kernel ridge regression}
In addition to RF regression, we also applied KRR to verify the robustness of our predictions. KRR implements the L2 regularization to prevent over-fitting of the data, and takes into account different nonlinear effects by invoking various kinds of kernel functions. Within KRR, the predicted target property, $f^{ML}(j)$, of the $j^{th}$ material is expressed as: $ f^{ML}(j) = \sum_{i}^{n}\alpha_i\mathcal{K}(\vec{x}_i,\vec{x}_j)$
Here, the summation is carried out over the entire training set with $n$ materials, $\alpha_i$'s are the weights and $\mathcal{K}(\vec{x}_i,\vec{x}_j)$ is the kernel function, representing the Euclidean distance between the feature vectors of the $i^{th}$ material from the training set and the $j^{th}$ material. To predict the formation energies of TMDs, we have used the Gaussian kernel, defined as: $\mathcal{K}(\vec{x}_i,\vec{x}_j) = exp(-\gamma|\vec{x}_i-\vec{x}_j|^2)$.
Training the KRR model involves finding the optimized values of the kernel coefficients, $\alpha_i$, and the Gaussian width, $\gamma$, through iterative minimization of the errors in the prediction.

\begin{figure*}[ht]
    \includegraphics[width=0.8\linewidth]{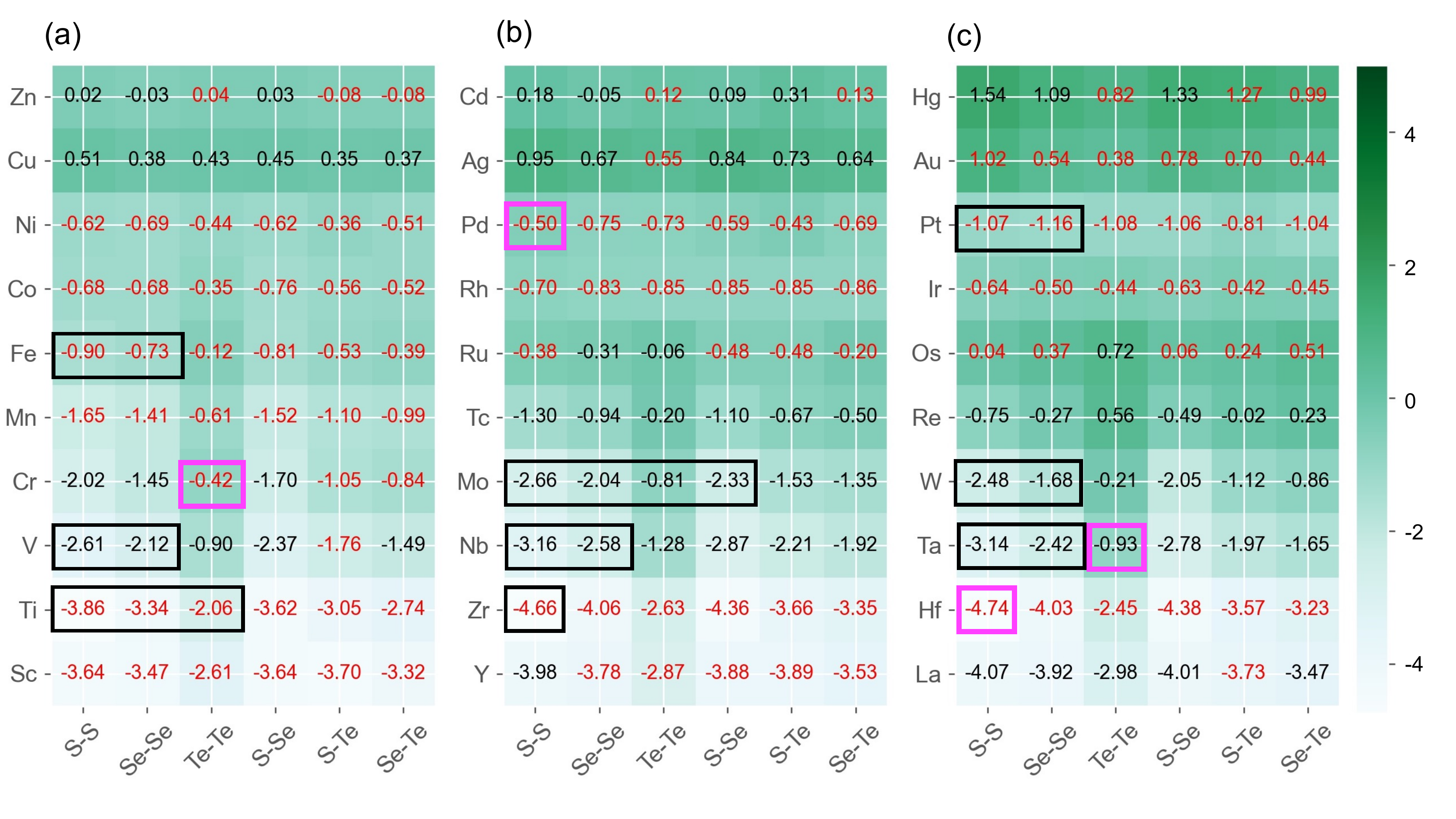}
    \caption{Heatmaps showing calculated formation energies (in eV per formula unit) of traditional and Janus TMD monolayers of (a) $3d$, (b) $4d$, and (c) $5d$-series elements in their most stable phase. The values printed in black (red) color correspond to the TMDs that prefer the \textit{1H} phase (\textit{1T} phase). Black boxes highlight examples of TMDs that have already been synthesized in either the \textit{1H} phase or \textit{1T} phase, whereas those in pink already exist in multilayers or in bulk, with a possibility of creating monolayers from those structures.}
    \label{fig:FEs}
\end{figure*}


\section{Results and discussion}



\subsection{Monolayer TMD Dataset}

The flowchart depicted in Fig.~\ref{fig:schematics}(d) outlines our systematic combinatorial search of different 2D TMD monolayers. Starting with 1080 total structures that consisted of FM, AFM and NM magnetic orderings for both \textit{1H} and \textit{1T} phases of the TMDs, we obtained the lowest energy structures from amongst the three possible magnetic orderings. This provided us with 180 structures within each of the two phases. These 360 structures were subsequently used to create machine learning models (detailed discussion in the next section). The pie chart in Fig.~\ref{fig:schematics}(e) shows the relative percentages of the structures that are FM, AFM or NM for the TMDs in the \textit{1H} and \textit{1T} phases. In both phases, most of the TMDs are nonmagnetic, consistent with previous results for already known TMDs~\cite{Ataca2012,Rasmussen2015}. Nevertheless, there is still a significant fraction of TMDs that adopt FM or AFM structures and can potentially have tremendous technological impact as 2D magnets. The pie charts in Fig.~\ref{fig:schematics}(e), also reveal that a larger percentage of structures in the \textit{1H} phase (as compared to those in \textit{1T} phase) adopted either FM or AFM ordering. 
The preferred magnetic orderings of each TMD in the \textit{1H} and \textit{1T} phases, along with their spin polarization energies and exchange energies per formula unit (f.u.\@) are provided in the supplement [see Supplementary Tables S1-S3 \cite{supp}]. Moreover, these results are in agreement with previous experimental and theoretical reports~\cite{Gao2013,Rasmussen2015,Zhou2018,Lu2019,OHara2018,Kanazawa2016,Mounet_Marzari2018}.


In order to study the chemical stability of the screened traditional and Janus TMDs, we computed the formation energies of the relaxed structures using Eqn.\@\,(\ref{eq1}). Further comparison between the two phases for each of the TMDs reduces the number of lowest energy structures to 180 (either \textit{1H} or \textit{1T} phase). Out of these 180 structures, 74 TMDs preferred to be in the \textit{1H} phase and 106 preferred to be in the \textit{1T} phase. Fig.~\ref{fig:FEs} shows the computed formation energies/f.u. of monolayers of the 180 TMDs in their most favored phase. The formation energy values given in black and red correspond to the TMD monolayers that preferentially form in the \textit{1H} phase and \textit{1T} phase, respectively. Some of these 180 TMDs have already been synthesized (examples indicated by black boxes in Fig.~\ref{fig:FEs}). TMDs encased in pink boxes in Fig.~\ref{fig:FEs} are examples of those materials for which bulk counterparts exist, with a possibility of creating the monolayers from those structures. 
We find that several TMDs exhibit small energy differences (less than 30\,meV/f.u.\@) between the \textit{1H} and \textit{1T} phases, and are therefore promising phase-change materials. These are: VTe$_2$, VSTe, VSeTe, ZnSe$_2$, YSe$_2$, YSSe, RuS$_2$, RuSe$_2$, TaTe$_2$, and OsTe$_2$ [see Supplementary Tables S1-S3 \cite{supp}]. The small energy differences suggest that these TMDs, can either exist in both phases (one of them being metastable), or can readily undergo a phase transition between the two phases upon application of external stimuli, such as gating, doping and applying strain.  Owing to a similar energy difference in two of its lowest energy phases, one of the well-known TMDs, MoTe$_2$ undergoes a phase transition from the \textit{1H} phase to the distorted octahedral symmetry, $T^\prime$ phase~\cite{Duerloo2014, Manchanda2020}.  The latter phase was not included in this study. Also note that a few of the TMDs, such as WTe$_2$, ReS$_2$ and ReSe$_2$, are known to form in low symmetry distorted phases -- $T^\prime$ and $T^{\prime \prime}$~\cite{Santosh2015,Zhou2018,Zhong2015}.  However, as the \textit{1H} and \textit{1T} phases are the most common phases among TMDs, we have not included the other possible structures at this stage. In addition, it is also possible for these materials to exist in higher symmetry phases (which may be metastable states for the crystals). Nevertheless, the dynamical and thermal stabilities as well as distances from the convex hull of the most promising novel TMDs, which are predicted to be chemically stable in this work [listed in Supplementary Tables S1-S3 \cite{supp}], need to be explored in future studies.

 Figures~\ref{fig:FEs}(a)-(c) show which of the TMDs are thermodynamically stable (i.e.\ have negative values for the formation energy). These formation energies become less negative across the TM-series (e.g., across the $3d$-series) and finally become positive, suggesting that the TMDs, which involve TM elements at the end of the series, are unstable. The inclusion of a large number of conventional and Janus TMDs in this work allowed us to effectively capture and recognize physicochemical trends in thermodynamically stable TMDs. The data trends in the calculated formation energies [Fig.~\ref{fig:FEs}] hint at two different connections. First, the number of d-electrons increases, while the atomic radii of the TM atoms generally decrease as we go across a period (e.g., from left to right in the $3d$-series of TM).  As mentioned above, the stability of the TMDs also decreases as the TM radii decrease across a period, implying that the TM radii and the d-electron counts play a major role in the stability of the structures. Second, for a fixed transition element (e.g., Mo), the formation energy decreases as one goes from S $\rightarrow$ Se $\rightarrow$ Te (for example, see the $\Delta E_f$s listed in Fig.~\ref{fig:FEs} for MoS$_2$, MoSe$_2$, and MoTe$_2$). This trend is generally seen for most of the stable traditional and Janus TMDs, though not for all. For a fixed TM, what has varied are the properties of the chalcogens, such as their electronegativities and their radii, which generally increase as one goes down a group. 

The aforementioned observations are consistent with Pauling's rules~\cite{Pauling1929} for ionic solids, and the work by F. R. Gamble~\cite{Gamble1974} on layered dichalcogenides, according to which the ratio of the atomic radii of the cations and anions, and the difference in their electronegativities play an important role in deciding the geometrical phase preferences of these materials. They are also consistent with the findings of Kertesz \textit{et al.}~\cite{Hoffmann_1984}, who explained the dependence of phase stability on d-electron counts. Hence, even before we apply ML algorithms, the observed trends hint at relationships between the formation energies and different attributes of the elements within the TMDs. Nevertheless, a deeper analysis is required to reveal and quantify the connections between different attributes of elements involved in a TMD, and its formation energy and phase stability.


\subsection{Down-selection of features}


\begin{figure*}[ht]
    \centering
    \includegraphics[width=0.8\linewidth]{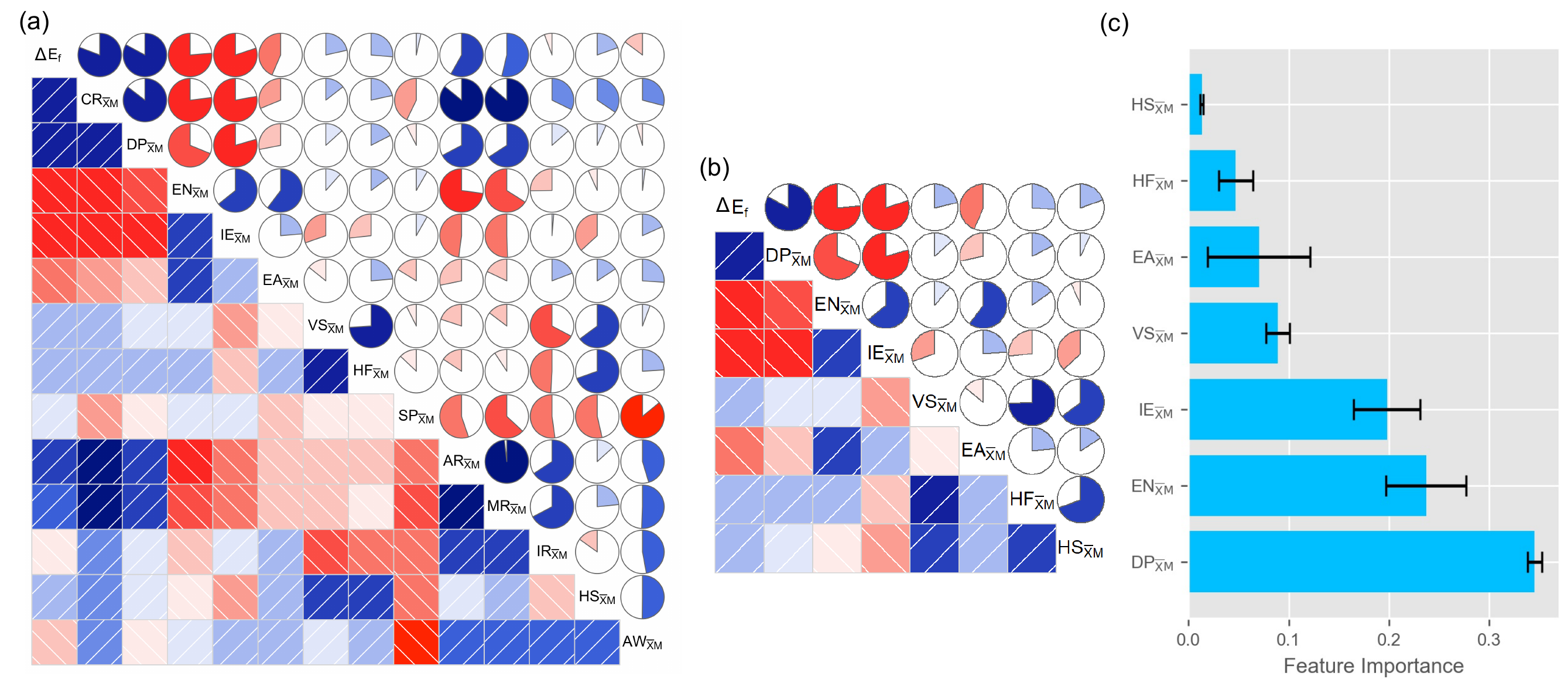}
    \caption{Down-selection of features: Pearson correlation for (a) all thirteen features from Set B and (b) seven down-selected features from Set B, summarizing the pairwise correlation of different features among themselves and with the target property vector (formation energy, $\Delta E_f$). The target property and the features are listed along the diagonal; the blue and red color correspond to positive and negative correlation coefficients, respectively. The boxes in the lower triangle and the pie charts in the upper triangle are two different pictorial representations of the same information, with the absolute value of the associated Pearson correlation coefficient given by either the lighter and darker shades of the colors in the boxes, or the filled fraction of the pie chart. (c) Relative importance of Set B feature vectors in predicting the formation energies of TMDs computed using the RF algorithm. The subscript $\bar{\textrm X}$M indicates that the down-selected compound features are obtained as differences of the average of attributes of the chalcogens with the that of the metal atom's attributes.}
    \label{fig:Corr_plot}
\end{figure*}


A large number of  descriptors increases the dimensionality and complexity of the machine learning model, not only making it difficult to interpret the results, but also leading to over-fitting. As there were a large number of compound descriptors (see the Technical Details section), we first needed to find a subset of highly relevant feature vectors in this large phase space of variables. 
To do so, we calculated the pairwise Pearson correlation coefficients for all feature vectors, as well as the correlation coefficients between the feature vectors and the target property vector (i.e., formation energy), which was calculated using DFT. If any two compound features showed $|\mathcal{P}| > 0.85$, only the one with the larger correlation with the target property was kept. Simultaneously, we also required the feature vectors to have $|\mathcal{P}| >0.15$ with the property vector.  We found that the feature vectors from Sets A and B not only have large pairwise correlations with each other, but also have comparable pairwise correlations with the target property [see Supplementary Table S4 \cite{supp}].  In addition, we found that compound feature vectors from Set C, created by subtracting the elemental attributes of the two chalcogen atoms (involved in a TMD), showed poor correlation with the target property and hence, were discarded. Therefore, in order to reduce the complexity of the model, we considered only the feature vectors of Set B.  

The pairwise correlations of the feature vectors from Set B amongst themselves and with the target property are shown in a correlogram chart in Fig.~\ref{fig:Corr_plot}(a). The absolute values of feature-feature and feature-target property correlations are pictorially represented through pie charts in the upper triangle and colored boxes in the lower triangle, both presenting the same information. The blue and red colors corresponds to positive and negative correlations, and the relative intensity of their colors corresponds to the relative strengths. Figure~\ref{fig:Corr_plot}(a) shows that not only are the covalent radii and dipole polarizabilities highly correlated, even their feature-target property correlations are comparable, being 0.81 for CR$_{\bar{\textrm X}\textrm M}$ and 0.83 for DP$_{\bar{\textrm X}\textrm M}$.  The large linear dependence between the covalent radii and dipole polarizabilities is understandable because the atomic dipole polarizability is highly correlated with an atom's size. The larger the size, the more polarizable an atom is, and in turn, the more readily it will form covalent bonds. It should also be pointed out that it is the differences in the covalent radii of the atoms involved that are more important as a feature as compared to the differences in ionic radii [see Fig.~\ref{fig:Corr_plot}(a)]. In contrast, in the earlier work by Gamble~\cite{Gamble1974}, wherein the author studied deciding factors that determine phase preferences (though not the formation energies), it was found that the ionic radii are more important attributes. However, TMDs are mostly covalent, with a very small fractional ionic character~\cite{Huisman1971,Gamble1974}. Our correlation analysis highlights this fact by assigning a much greater correlation fraction to covalent radii differences as compared to the ionic radii differences. As the covalent radii and dipole polarizabilities show comparable feature-target property correlations, we decided to create two different subsets of down-selected features based on our aforementioned criteria -- one subset included DP$_{\bar{\textrm X}\textrm M}$ and another included CR$_{\bar{\textrm X}\textrm M}$. From this point forward, we present our analysis for the short-listed feature subset (from set B) that includes DP$_{\bar{\textrm X}\textrm M}$, while the analysis for the CR$_{\bar{\textrm X}\textrm M}$\,-containing subset is presented in Supplementary Note II \cite{supp}.

\begin{figure*}[ht]
    \centering
    \includegraphics[width=0.8\linewidth]{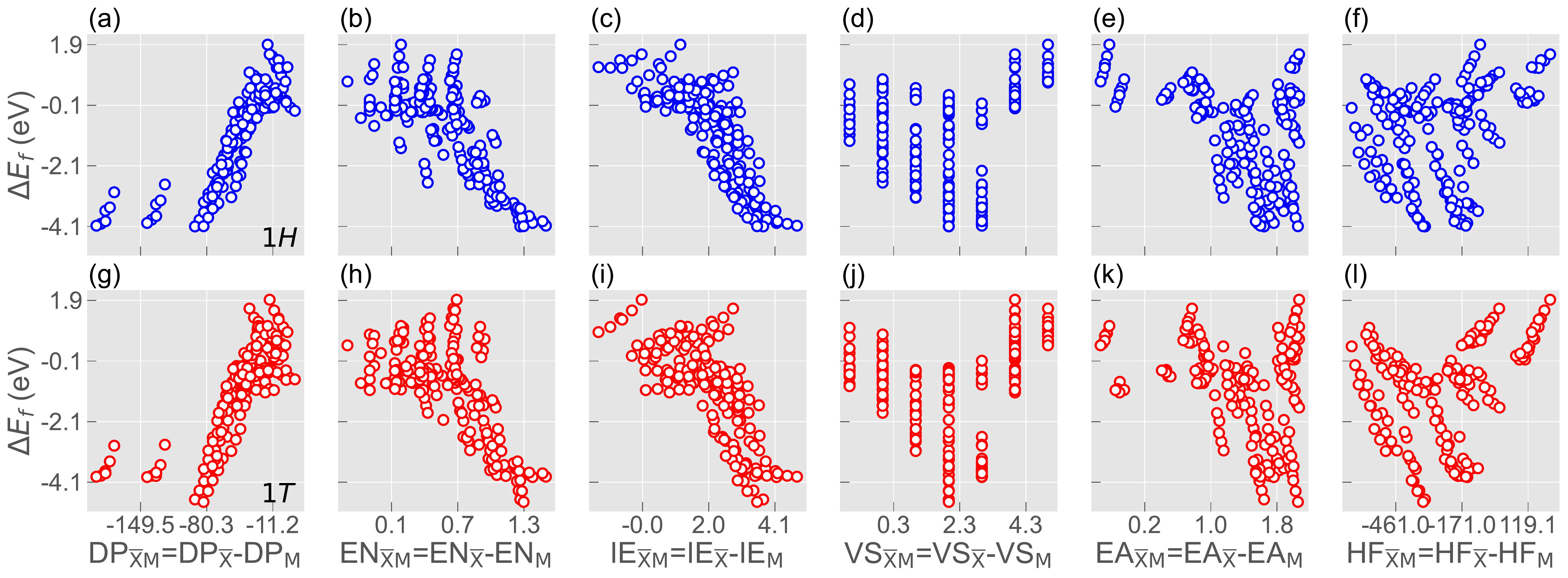}
    \
    \hspace{1cm}
    \caption{Formation energy vs. the down-selected feature vectors for different TMDs in: (a-f) the \textit{1H} phase and (g-l) \textit{1T} phase. Along the x-axis, the compound feature vectors are computed from elemental features according to: $\mathcal{F}_{\bar{\textrm X}\textrm M}=\frac{1}{2} (\mathcal{F}_{\textrm X}+\mathcal{F}_{\textrm Y}) - \mathcal{F}_{\textrm M}$.
}
    \label{fig:features}
\end{figure*}

The correlogram chart in Fig.~\ref{fig:Corr_plot}(b) summarizes the pairwise correlations between the DP$_{\bar{\textrm X}\textrm M}$-containing subset of seven down-selected features. Thus far, we have used Pearson correlation for the first round of feature down-selection, which helped to reduce the dimensionality of the problem. 
The next round of down-selection of features from Set B came from the ML algorithms themselves. In this round of down-selection, we calculated the relative feature importances using an RF model, aiming to predict the thermodynamic stability of the TMDs.
As mentioned in the previous section, the RF regression algorithm has two main hyperparameters: number of trees and maximum allowable depth. Tuning these parameters is essential to avoid sub-optimal performance and overfitting of the model. Therefore, to find optimal values of these parameters, we trained the RF regression algorithm over 90\% of the dataset and tested the model with 10\% unseen data for each set of hyperparameters.  The ML model was trained over 20 randomly distributed train/test splits of data for different k-fold cross-validations (with $\mathrm{k}=5, 8, 10$).
 We then computed the average root mean squared error (RMSE) and R$^2$-score (goodness-of-fit score) using the dataset from the \textit{1H} phase of the TMDs.  
The average R$^2$-scores and RMSEs for different cross-validations are given in the Supplementary Table S5 \cite{supp}. A 10-fold cross-validation was used in the rest of the work as it shows very good training and testing R$^2$-scores and RMSE-values. The optimized number of trees and maximum allowable depth for 10-fold cross-validation are 48 and 16, respectively.  Using the optimized values of hyperparameters, we computed the importance of different features in predicting the formation energies of the TMDs. 
The bar graph in Fig.~\ref{fig:Corr_plot}(c) shows the mean values of the relative feature importance, with the error bars representing their standard deviations, obtained from 20 randomly distributed train/test splits of data.  Fig~\ref{fig:Corr_plot}(c) shows that DP$_{\bar{\textrm X}\textrm M}$, EN$_{\bar{\textrm X}\textrm M}$, and IE$_{\bar{\textrm X}\textrm M}$ are the most important features, with a relatively small contribution from VS$_{\bar{\textrm X}\textrm M}$, EA$_{\bar{\textrm X}\textrm M}$ and HF$_{\bar{\textrm X}\textrm M}$, corroborating the results from our pair-wise correlation study. Since HS$_{\bar{\textrm X}\textrm M}$ has a negligible effect on the prediction accuracy of the formation energy, following the general \textit{principle of parsimony}, we excluded HS$_{\bar{\textrm X}\textrm M}$ from the down-selected feature set. The six chosen features were used to design machine learning models.

Before implementing the ML algorithms, we would like to further elaborate upon the physically-intuitive results that emerge from our down-selection analysis of the feature set. 
In Fig.~\ref{fig:features}, we plot the down-selected features as a function of the formation energy for the \textit{1H} [Figs~\ref{fig:features}(a)--(f)] and \textit{1T} [Figs~\ref{fig:features}(g)--(l)] phases.  Overall, one can see a strong dependence of formation energy on DP$_{\bar{\textrm X}\textrm M}$ [equivalently, CR$_{\bar{\textrm X}\textrm M}$ in Supplementary Fig. S1 \cite{supp}], EN$_{\bar{\textrm X}\textrm M}$, IE$_{\bar{\textrm X}\textrm M}$, and VS$_{\bar{\textrm X}\textrm M}$, making them the key features. 
 Figures~\ref{fig:features}(a) and (g) show that the data are more scattered as the difference in dipole polarizabilities (DP$_{\bar{\textrm X}\textrm M}$)  decreases [same trend for covalent radii], indicating that dipole polarizabilities (covalent radii) alone are not a deciding factor in the thermodynamical stability of TMDs. Another important key feature vector is the difference in electronegativities of the TM atom and chalcogens (EN$_{\bar{\textrm X}\textrm M}$) which accounts for the fractional ionic character [Figs.~\ref{fig:features}(b) and (h)]. The  fractional ionic character, $f_{i}$, as defined by Pauling is given by: $f_{i}=1-exp[-(\mathrm{EN}_{A} -\mathrm{EN}_{B})^{2}/4]$, where $EN_{A}$ and $EN_{B}$ are the respective elemental electronegativities of a binary compound of elements A and B. Although there is a spread of the formation energy values, the overall trend shows a decrease in the formation energy with increasing values of EN$_{\bar{\textrm X}\textrm M}$. The relative ionization energies of atoms forming the materials (IE$_{\bar{\textrm X}\textrm M}$) is generally related to how readily the respective atoms enter into chemical reactions and form bonds with each other. It is, therefore, not surprising that this is one of the important features selected by the correlation analysis [Figs~\ref{fig:features}(c) and (i)].  Lastly, the relative difference in valencies of the cations and anions,VS$_{\bar{\textrm X}\textrm M}$, shows a correlation with the formation energies, displaying a U-shaped trend, with the bottom of the U-shape at $\mathrm{V}_{\bar{\textrm X}\textrm M}=2$ [Figs~\ref{fig:features}(d) and (j)].  In the compound attribute, VS$_{\bar{\textrm X}\textrm M}$, the average valency of the chalcogens is fixed as we go from one TMD to the next, while that of the transition element changes according to its d-electron count.  This trend for VS$_{\bar{\textrm X}\textrm M}$ is understandable, based on work by Kertesz \textit{et al.}~\cite{Hoffmann_1984}, who explained how the stability of TMDs changes as a function of d-electron counts.  Although no such obvious trends can be seen for the EA$_{\bar{\textrm X}{\textrm M}}$  [Figs.~\ref{fig:features}(e) and (k)] and HF$_{\bar{\textrm X}{\textrm M}}$  [Figs.~\ref{fig:features}(f) and (l)] feature vectors, they were included when creating an ML regression model because the calculated feature importance [Fig.~\ref{fig:Corr_plot}(b)] for these two attributes is comparable to that for the valence state. 

\begin{figure*}[ht]
    \includegraphics[width=0.85\linewidth]{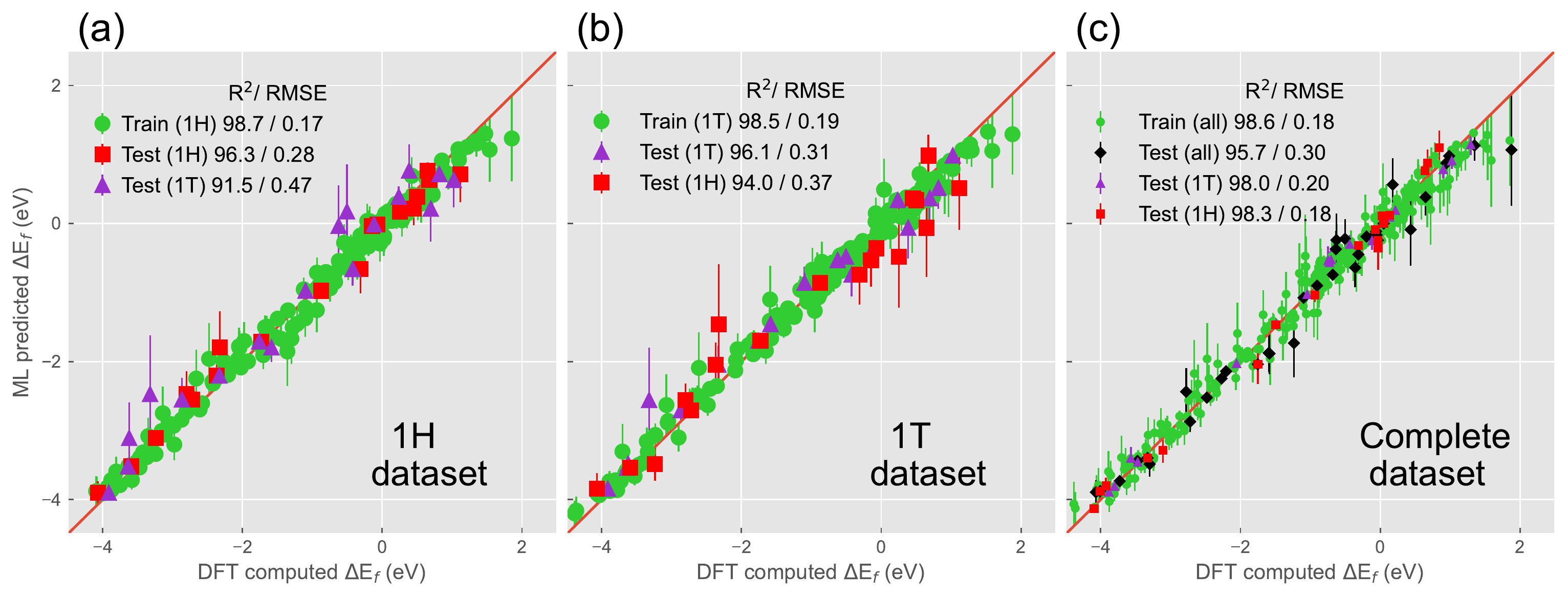}
    \caption{Parity plots of formation energies of: (a) \textit{1H}, (b) \textit{1T} phases of TMD monolayers, and (c) the entire dataset. The DFT computed formation energies are along the abscissa and those predicted using RF along the ordinate. Each dataset is divided such that 90\% of the data is used to train the RF regression model and 10\% data is used for validation of the model. The error bars on each data point correspond to prediction errors.}
    \label{fig:ML_figure}
\end{figure*}


\subsection{Machine learning algorithms for regression and classification}  

\subsubsection{Regression algorithms for formation energies}
Having established the physical rationale behind the down-selected feature vectors, we applied regression algorithms to predict the formation energy of TMDs. The robustness of the statistical models was tested in several steps, which are described below. 
Unless stated otherwise, all ML regression models were trained on \textit{1H}, \textit{1T} or the complete datasets using 20 different randomly selected train/test splits of 90\%/10\%. 
The resulting average R$^2$-scores and RMSE values were then calculated to determine each model's performance. 


We first trained the RF regression model to predict formation energies of TMDs in the \textit{1H} phase. 
With the six down-selected features, the average training and test R$^2$-scores are 98.7\% and 96.3\%, respectively. The corresponding average RMSEs for the training and test data points are 0.17\,eV and 0.28\,eV, respectively. The prediction performance of the model for the \textit{1H} dataset is shown in Fig.~\ref{fig:ML_figure}(a). Most of the data points lie on or are near the straight line (so-called line of accuracy), suggesting that the model is highly accurate. The deviation from the line of accuracy is quantified by calculating the error in predicting the DFT computed formation energy at each point and is indicated by an error bar. The error bars for the data points with negative formation energies are relatively smaller than those for data points with positive formation energies, indicating that our trained RF regression model is able to accurately predict systems that are structurally stable. We also went one step further and tested the prediction performance of our ML model trained on the \textit{1H} data and testing it with 10\% of the \textit{1T} phase dataset, with accuracy scores of 91.5\%.\newline


Fig.~\ref{fig:ML_figure}(b) plots the prediction performance of the RF model trained on the \textit{1T} dataset.
This model achieves an average accuracy of 98.5\% for training data and 96.1\% for test data, with the corresponding RMSEs of 0.19\,eV and 0.31\,eV respectively.  The prediction performance of our ML model trained on the \textit{1T} data was also tested on 10\% of the \textit{1H}-phase dataset, with an accuracy score of 94.0\%. The average RMSEs numbers for the \textit{1H} and \textit{1T} testing datasets are comparable [see Figs~\ref{fig:ML_figure}(a) and (b)], showing the robustness of the ML model in predicting the chemical stability of both crystal geometries. Hence, these results show that the six-feature ML model is efficient in predicting the formation energy over a vast chemical space and for different phases. Furthermore, we verified the robustness of the ML model by showing that our results are insensitive to the train/test split [see Supplementary Tables S6-S7 \cite{supp}]. 


So far, the ML model focussed on predicting formation energies of TMDs in either the \textit{1H} or \textit{1T} phase. We further checked the prediction performance of our model by applying a more stringent test.  We considered the entire dataset of 1080 candidate structures and augmented the feature set by four additional feature vectors (one feature vector each to indicate FM, AFM or NM ground state, and one feature vector corresponding to the phase). We then trained the RF regression model with the set of hyperparameters that were optimized using the \textit{1H} dataset alone. 
The calculated R$^2$-score and RMSEs are shown in Fig.~\ref{fig:ML_figure}(c). The prediction performance of the enhanced ML model across a different dataset suggests that it can be further generalized to other TMDs that preferentially form configurations different from either the \textit{1H} or \textit{1T} phases. 

In order to check the validity of our results over different algorithms, we also applied KRR to the \textit{1H} dataset. The key hyperparameters of the KRR model, namely, $\alpha$ and $\gamma$ were obtained through iterative minimization of the errors in the prediction. In our case, we calculated these hyperparameters by training the KRR model over 20 randomly distributed train/test splits of data in the ratio of 90\%/10\% and minimizing the root mean squared error (RMSE) of the model. 
In order to avoid overfitting in the model, we used a 10-fold cross validation. Using this scheme, we calculated the optimized values of $\alpha$ and $\gamma$ to be 0.0001 and 0.0001, respectively.  We then separately trained KRR model on the \textit{1H} and \textit{1T} datasets using the optimized hyperparameters of KRR model.
For the \textit{1H} dataset, we calculated the average R$^2$-score and RMSE over 20 random train/test data splits to be 98.7\%/95.3\% and 0.17/0.30\,eV, respectively. 
The average R$^2$-score and RMSE for the \textit{1T} phase were 99.1\%/96.3\% and 0.14/0.28\,eV, respectively. The parity plots of these trained datasets using KRR algorithm are shown in Supplementary Fig. S5 \cite{supp}. From Figs~\ref{fig:ML_figure} and S5, we can see that the corresponding accuracies of the RF regression and the KRR models are comparable, proving that our down-selected features are robust across different algorithms and describe the material properties of TMDs with a fairly high degree of accuracy.\newline

\begin{figure*}[ht]
    \includegraphics[width=0.75\linewidth]{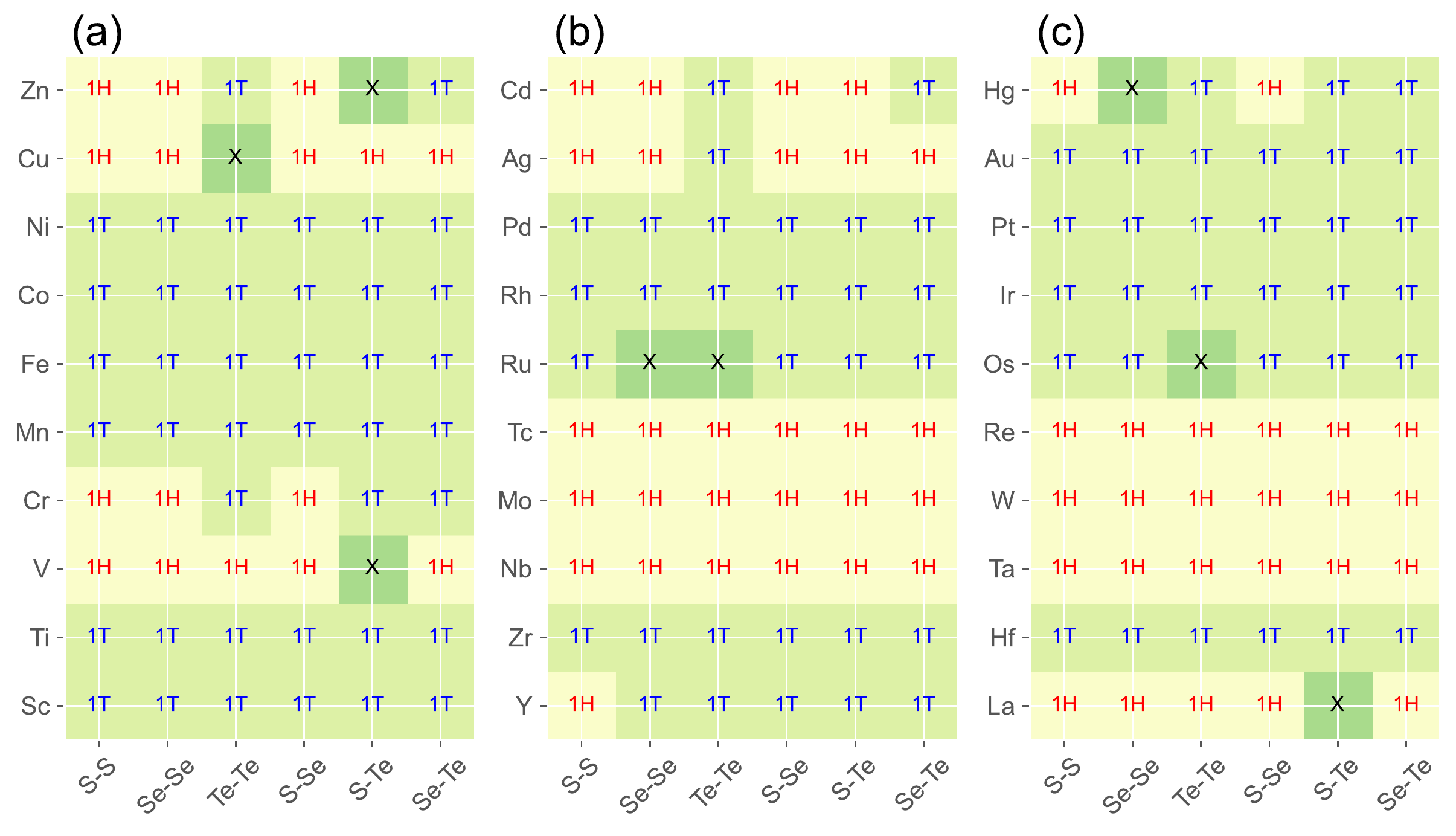}
    \caption{Phase-preference prediction performance of RF classification algorithm for the 180 down selected TMDs (see Fig.~\ref{fig:FEs}). The boxes marked with an $``$X$"$ correspond to systems that were not predicted accurately by our model. We consider our prediction to be accurate if we obtain the correct phase at least 90\% of the time.}
    \label{fig:phase_pred}
\end{figure*}

\subsubsection{Random Forest classification algorithm for predicting phase preference}

So far, we have shown that we have built an efficient and accurate machine learning model to predict the formation energies of different TMDs using six descriptors. We next tested our model to see if it can predict the phase preferences of TMDs [see Fig.~\ref{fig:FEs}]. Starting with the down selected features and the hyperparameters from the RF regression model discussed above, we augmented the feature set by the addition of one categorical feature to indicate the preferred phases of the TMDs. We then trained the RF classification model and used it to predict the phase preferences of the TMDs, classifying them as \textit{1H}- and \textit{1T} phase crystals.  In order to avoid overfitting with respect to the randomly selected data, and to ensure that our results were converged, we trained the model over different numbers of random train/test splits of data, predicting the unseen data each time. We found that the  correctly predicted phases do not change if we use more than 55 different random train/test splits of data. This analysis yielded a high training and test score of 98.6\% and 91.8\%, respectively. The preferred phases predicted for the 180 down-selected traditional and Janus structures are shown in Fig.~\ref{fig:phase_pred}. We were able to accurately predict the phases of 172 monolayers. We note that for a small fraction of materials, the RF classification model did not predict the phase preference accurately; these are marked with an ``X" in the figure. A comparison of Figs~\ref{fig:FEs} and~\ref{fig:phase_pred} shows that some of the TMDs for which our model could not predict the phases accurately are the chemically unstable ones (positive formation energies). However, some of the chemically stable TMDs are also predicted inaccurately (e.g. RuSe$_2$, RuTe$_2$, and LaSeTe). There may be several reasons for these inaccuracies. Some of these systems either have a relatively smaller differences in energies between the \textit{1H} and \textit{1T} phases, or they may prefer other distorted phases (\textit{1T}$'$ or ZT), which were not considered in this study. 




It is interesting to note that the same set of down-selected compound features account for both the chemical stability and the relative phase stability of the TMDs. This implies that the RF regression algorithm itself can be used to predict the correct phase ordering by determining the differences between predicted formation energies for \textit{1H}- and \textit{1T}-phase TMDs. Supplementary Fig. S6 \cite{supp} shows that the RF regression model indeed predicts the formation energies for the two phases at the level of accuracy needed to predict phase-preferences for the TMDs, correctly predicting the crystal structures of 169 out of 180 TMDs.


\section{Conclusion}
%

Predicting phase-preferences of TMDs from just the knowledge of the constituent atoms is important, but has remained a long-standing problem. Earlier works  had used packing fraction considerations (starting with Pauling~\cite{Pauling1929} for ionic crystals), an empirical measure consisting of ionic radius ratios and degree of ionicity (Gamble~\cite{Gamble1974}), and/or electronic factors (d-electron counts by Kertesz \textit{et al.}~\cite{Hoffmann_1984}) to rationalize the observed structures of the traditional TMDs. Our machine learning-based quantitative analysis not only rediscovers each of these different physicochemical factors, but also uncovers previously unknown relative attributes of the constituent atoms that govern chemical stability and crystal structures of TMDs. 
Establishing different connections between compound attributes of the TMDs and the target property/properties enabled us to not only rationalize, but also to accurately predict phases for  171 out of 180 TMDs. 
The framework developed here can be generalized for other classes of materials, and provides a path forward in materials research.

\vspace{12pt}

\begin{acknowledgments}

We acknowledge support by the National Science Foundation under NSF grant number DMR-1752840.  This work used the Extreme Science and Engineering Discovery Environment (XSEDE) [under Project numbers PHY180014 and TG-DMR200008], which is supported by National Science Foundation Grant Number ACI-1053575. We also acknowledge the use of Mach cluster maintained by the Laboratory of Physical Sciences, Maryland. V.S. is supported by the Infrastructure for Scientific Applications and Advanced Computing (ISAAC) at the University of Tennessee. P.D. and V.S. acknowledge XSEDE Extended Collaborative Support Service (ECSS) program.  

\end{acknowledgments}

\vspace{12pt}
\noindent \textbf{\large{Author Contributions}}
\vspace{8pt}

\noindent  P.D. conceived and directed the study. P.K. and V.S. made equal contributions to the work and performed the DFT calculations.  P.K. wrote the ML codes, under the supervision of P.D. and V.S.  S.S. performed additional searches for compound descriptors to help with the revisions. All authors were involved in the analysis of results and discussions.  P.D. wrote the final manuscript, with inputs from the initial draft from P.K. All authors proof-read and reviewed the final revision of the manuscript.

\normalem

\bibliographystyle{apsrev}


\begin{thebibliography}{61}

\makeatletter
\providecommand \@ifxundefined [1]{%
 \@ifx{#1\undefined}
}%
\providecommand \@ifnum [1]{%
 \ifnum #1\expandafter \@firstoftwo
 \else \expandafter \@secondoftwo
 \fi
}%
\providecommand \@ifx [1]{%
 \ifx #1\expandafter \@firstoftwo
 \else \expandafter \@secondoftwo
 \fi
}%
\providecommand \natexlab [1]{#1}%
\providecommand \enquote  [1]{``#1''}%
\providecommand \bibnamefont  [1]{#1}%
\providecommand \bibfnamefont [1]{#1}%
\providecommand \citenamefont [1]{#1}%
\providecommand \href@noop [0]{\@secondoftwo}%
\providecommand \href [0]{\begingroup \@sanitize@url \@href}%
\providecommand \@href[1]{\@@startlink{#1}\@@href}%
\providecommand \@@href[1]{\endgroup#1\@@endlink}%
\providecommand \@sanitize@url [0]{\catcode `\\12\catcode `\$12\catcode
  `\&12\catcode `\#12\catcode `\^12\catcode `\_12\catcode `\%12\relax}%
\providecommand \@@startlink[1]{}%
\providecommand \@@endlink[0]{}%
\providecommand \url  [0]{\begingroup\@sanitize@url \@url }%
\providecommand \@url [1]{\endgroup\@href {#1}{\urlprefix }}%
\providecommand \urlprefix  [0]{URL }%
\providecommand \Eprint [0]{\href }%
\providecommand \doibase [0]{http://dx.doi.org/}%
\providecommand \selectlanguage [0]{\@gobble}%
\providecommand \bibinfo  [0]{\@secondoftwo}%
\providecommand \bibfield  [0]{\@secondoftwo}%
\providecommand \translation [1]{[#1]}%
\providecommand \BibitemOpen [0]{}%
\providecommand \bibitemStop [0]{}%
\providecommand \bibitemNoStop [0]{.\EOS\space}%
\providecommand \EOS [0]{\spacefactor3000\relax}%
\providecommand \BibitemShut  [1]{\csname bibitem#1\endcsname}%
\let\auto@bib@innerbib\@empty

\bibitem[{\citenamefont{Wilson and Yoffe}(1969)}]{Wilson_Yoffe_1969}
\bibinfo{author}{\bibfnamefont{J.}~\bibnamefont{Wilson}} \bibnamefont{and}
  \bibinfo{author}{\bibfnamefont{A.}~\bibnamefont{Yoffe}},
  \bibinfo{journal}{Advances in Physics} \textbf{\bibinfo{volume}{18}},
  \bibinfo{pages}{193} (\bibinfo{year}{1969}),
  \urlprefix\url{https://doi.org/10.1080/00018736900101307}.

\bibitem[{\citenamefont{Yoffe}(1973)}]{Yoffe_review1973}
\bibinfo{author}{\bibfnamefont{A.~D.} \bibnamefont{Yoffe}},
  \bibinfo{journal}{Annual Review of Materials Science}
  \textbf{\bibinfo{volume}{3}}, \bibinfo{pages}{147} (\bibinfo{year}{1973}),
  \urlprefix\url{https://doi.org/10.1146/annurev.ms.03.080173.001051}.

\bibitem[{\citenamefont{Gamble}(1974)}]{Gamble1974}
\bibinfo{author}{\bibfnamefont{F.}~\bibnamefont{Gamble}},
  \bibinfo{journal}{Journal of Solid State Chemistry}
  \textbf{\bibinfo{volume}{9}}, \bibinfo{pages}{358} (\bibinfo{year}{1974}),
  ISSN \bibinfo{issn}{0022-4596},
  \urlprefix\url{https://www.sciencedirect.com/science/article/pii/0022459674900954}.

\bibitem[{\citenamefont{Kertesz and Hoffmann}(1984)}]{Hoffmann_1984}
\bibinfo{author}{\bibfnamefont{M.}~\bibnamefont{Kertesz}} \bibnamefont{and}
  \bibinfo{author}{\bibfnamefont{R.}~\bibnamefont{Hoffmann}},
  \bibinfo{journal}{Journal of the American Chemical Society}
  \textbf{\bibinfo{volume}{106}}, \bibinfo{pages}{3453} (\bibinfo{year}{1984}),
  \urlprefix\url{https://doi.org/10.1021/ja00324a012}.

\bibitem[{\citenamefont{Novoselov et~al.}(2005)\citenamefont{Novoselov, Jiang,
  Schedin, Booth, Khotkevich, Morozov, and Geim}}]{Novoselov2005}
\bibinfo{author}{\bibfnamefont{K.~S.} \bibnamefont{Novoselov}},
  \bibinfo{author}{\bibfnamefont{D.}~\bibnamefont{Jiang}},
  \bibinfo{author}{\bibfnamefont{F.}~\bibnamefont{Schedin}},
  \bibinfo{author}{\bibfnamefont{T.~J.} \bibnamefont{Booth}},
  \bibinfo{author}{\bibfnamefont{V.~V.} \bibnamefont{Khotkevich}},
  \bibinfo{author}{\bibfnamefont{S.~V.} \bibnamefont{Morozov}},
  \bibnamefont{and} \bibinfo{author}{\bibfnamefont{A.~K.} \bibnamefont{Geim}},
  \bibinfo{journal}{Proceedings of the National Academy of Sciences}
  \textbf{\bibinfo{volume}{102}}, \bibinfo{pages}{10451}
  (\bibinfo{year}{2005}), ISSN \bibinfo{issn}{0027-8424},
  \urlprefix\url{https://www.pnas.org/content/102/30/10451}.

\bibitem[{\citenamefont{Ayari et~al.}(2007)\citenamefont{Ayari, Cobas,
  Ogundadegbe, and Fuhrer}}]{Ayari_2007_JAP}
\bibinfo{author}{\bibfnamefont{A.}~\bibnamefont{Ayari}},
  \bibinfo{author}{\bibfnamefont{E.}~\bibnamefont{Cobas}},
  \bibinfo{author}{\bibfnamefont{O.}~\bibnamefont{Ogundadegbe}},
  \bibnamefont{and} \bibinfo{author}{\bibfnamefont{M.~S.}
  \bibnamefont{Fuhrer}}, \bibinfo{journal}{Journal of Applied Physics}
  \textbf{\bibinfo{volume}{101}}, \bibinfo{pages}{014507}
  (\bibinfo{year}{2007}), \urlprefix\url{https://doi.org/10.1063/1.2407388}.

\bibitem[{\citenamefont{Ramakrishna~Matte
  et~al.}(2010)\citenamefont{Ramakrishna~Matte, Gomathi, Manna, Late, Datta,
  Pati, and Rao}}]{Matte_2010}
\bibinfo{author}{\bibfnamefont{H.}~\bibnamefont{Ramakrishna~Matte}},
  \bibinfo{author}{\bibfnamefont{A.}~\bibnamefont{Gomathi}},
  \bibinfo{author}{\bibfnamefont{A.}~\bibnamefont{Manna}},
  \bibinfo{author}{\bibfnamefont{D.}~\bibnamefont{Late}},
  \bibinfo{author}{\bibfnamefont{R.}~\bibnamefont{Datta}},
  \bibinfo{author}{\bibfnamefont{S.}~\bibnamefont{Pati}}, \bibnamefont{and}
  \bibinfo{author}{\bibfnamefont{C.}~\bibnamefont{Rao}},
  \bibinfo{journal}{Angewandte Chemie International Edition}
  \textbf{\bibinfo{volume}{49}}, \bibinfo{pages}{4059} (\bibinfo{year}{2010}),
  \urlprefix\url{https://onlinelibrary.wiley.com/doi/abs/10.1002/anie.201000009}.

\bibitem[{\citenamefont{Coleman et~al.}(2011)\citenamefont{Coleman, Lotya,
  {\O'}Neill, Bergin, King, Khan, Young, Gaucher, De, Smith
  et~al.}}]{Coleman2011}
\bibinfo{author}{\bibfnamefont{J.~N.} \bibnamefont{Coleman}},
  \bibinfo{author}{\bibfnamefont{M.}~\bibnamefont{Lotya}},
  \bibinfo{author}{\bibfnamefont{A.}~\bibnamefont{{\O'}Neill}},
  \bibinfo{author}{\bibfnamefont{S.~D.} \bibnamefont{Bergin}},
  \bibinfo{author}{\bibfnamefont{P.~J.} \bibnamefont{King}},
  \bibinfo{author}{\bibfnamefont{U.}~\bibnamefont{Khan}},
  \bibinfo{author}{\bibfnamefont{K.}~\bibnamefont{Young}},
  \bibinfo{author}{\bibfnamefont{A.}~\bibnamefont{Gaucher}},
  \bibinfo{author}{\bibfnamefont{S.}~\bibnamefont{De}},
  \bibinfo{author}{\bibfnamefont{R.~J.} \bibnamefont{Smith}},
  \bibnamefont{et~al.}, \bibinfo{journal}{Science}
  \textbf{\bibinfo{volume}{331}}, \bibinfo{pages}{568} (\bibinfo{year}{2011}),
  \urlprefix\url{https://www.science.org/doi/abs/10.1126/science.1194975}.

\bibitem[{\citenamefont{Liu et~al.}(2012)\citenamefont{Liu, Zhang, Lee, Lin,
  Chang, Su, Chang, Li, Shi, Zhang et~al.}}]{Liu_MoS2_2012}
\bibinfo{author}{\bibfnamefont{K.-K.} \bibnamefont{Liu}},
  \bibinfo{author}{\bibfnamefont{W.}~\bibnamefont{Zhang}},
  \bibinfo{author}{\bibfnamefont{Y.-H.} \bibnamefont{Lee}},
  \bibinfo{author}{\bibfnamefont{Y.-C.} \bibnamefont{Lin}},
  \bibinfo{author}{\bibfnamefont{M.-T.} \bibnamefont{Chang}},
  \bibinfo{author}{\bibfnamefont{C.-Y.} \bibnamefont{Su}},
  \bibinfo{author}{\bibfnamefont{C.-S.} \bibnamefont{Chang}},
  \bibinfo{author}{\bibfnamefont{H.}~\bibnamefont{Li}},
  \bibinfo{author}{\bibfnamefont{Y.}~\bibnamefont{Shi}},
  \bibinfo{author}{\bibfnamefont{H.}~\bibnamefont{Zhang}},
  \bibnamefont{et~al.}, \bibinfo{journal}{Nano Letters}
  \textbf{\bibinfo{volume}{12}}, \bibinfo{pages}{1538} (\bibinfo{year}{2012}),
  \bibinfo{note}{pMID: 22369470},
  \urlprefix\url{https://doi.org/10.1021/nl2043612}.

\bibitem[{\citenamefont{Nakano et~al.}(2017)\citenamefont{Nakano, Wang,
  Kashiwabara, Matsuoka, and Iwasa}}]{Nakano_WSe2_2017}
\bibinfo{author}{\bibfnamefont{M.}~\bibnamefont{Nakano}},
  \bibinfo{author}{\bibfnamefont{Y.}~\bibnamefont{Wang}},
  \bibinfo{author}{\bibfnamefont{Y.}~\bibnamefont{Kashiwabara}},
  \bibinfo{author}{\bibfnamefont{H.}~\bibnamefont{Matsuoka}}, \bibnamefont{and}
  \bibinfo{author}{\bibfnamefont{Y.}~\bibnamefont{Iwasa}},
  \bibinfo{journal}{Nano Letters} \textbf{\bibinfo{volume}{17}},
  \bibinfo{pages}{5595} (\bibinfo{year}{2017}), \bibinfo{note}{pMID: 28849935},
  \urlprefix\url{https://doi.org/10.1021/acs.nanolett.7b02420}.

\bibitem[{\citenamefont{Kaur et~al.}(2018)\citenamefont{Kaur, Yadav,
  Srivastava, Singh, Rath, Schneider, Sinha, and Srivastava}}]{Kaur2018}
\bibinfo{author}{\bibfnamefont{H.}~\bibnamefont{Kaur}},
  \bibinfo{author}{\bibfnamefont{S.}~\bibnamefont{Yadav}},
  \bibinfo{author}{\bibfnamefont{A.~K.} \bibnamefont{Srivastava}},
  \bibinfo{author}{\bibfnamefont{N.}~\bibnamefont{Singh}},
  \bibinfo{author}{\bibfnamefont{S.}~\bibnamefont{Rath}},
  \bibinfo{author}{\bibfnamefont{J.~J.} \bibnamefont{Schneider}},
  \bibinfo{author}{\bibfnamefont{O.~P.} \bibnamefont{Sinha}}, \bibnamefont{and}
  \bibinfo{author}{\bibfnamefont{R.}~\bibnamefont{Srivastava}},
  \bibinfo{journal}{Nano Research} \textbf{\bibinfo{volume}{11}},
  \bibinfo{pages}{343} (\bibinfo{year}{2018}), ISSN \bibinfo{issn}{1998-0000},
  \urlprefix\url{https://doi.org/10.1007/s12274-017-1636-x}.

\bibitem[{\citenamefont{Puthirath et~al.}(2021)\citenamefont{Puthirath, Balan,
  Oliveira, Sreepal, Robles~Hernandez, Gao, Chakingal, Sassi, Thibeorchews,
  Costin et~al.}}]{Puthirath_JPCC_2021}
\bibinfo{author}{\bibfnamefont{A.~B.} \bibnamefont{Puthirath}},
  \bibinfo{author}{\bibfnamefont{A.~P.} \bibnamefont{Balan}},
  \bibinfo{author}{\bibfnamefont{E.~F.} \bibnamefont{Oliveira}},
  \bibinfo{author}{\bibfnamefont{V.}~\bibnamefont{Sreepal}},
  \bibinfo{author}{\bibfnamefont{F.~C.} \bibnamefont{Robles~Hernandez}},
  \bibinfo{author}{\bibfnamefont{G.}~\bibnamefont{Gao}},
  \bibinfo{author}{\bibfnamefont{N.}~\bibnamefont{Chakingal}},
  \bibinfo{author}{\bibfnamefont{L.~M.} \bibnamefont{Sassi}},
  \bibinfo{author}{\bibfnamefont{P.}~\bibnamefont{Thibeorchews}},
  \bibinfo{author}{\bibfnamefont{G.}~\bibnamefont{Costin}},
  \bibnamefont{et~al.}, \bibinfo{journal}{The Journal of Physical Chemistry C}
  \textbf{\bibinfo{volume}{125}}, \bibinfo{pages}{18927}
  (\bibinfo{year}{2021}),
  \urlprefix\url{https://doi.org/10.1021/acs.jpcc.1c04977}.

\bibitem[{\citenamefont{Chhowalla et~al.}(2013)\citenamefont{Chhowalla, Shin,
  Eda, Li, Loh, and Zhang}}]{Chhowalla2013}
\bibinfo{author}{\bibfnamefont{M.}~\bibnamefont{Chhowalla}},
  \bibinfo{author}{\bibfnamefont{H.~S.} \bibnamefont{Shin}},
  \bibinfo{author}{\bibfnamefont{G.}~\bibnamefont{Eda}},
  \bibinfo{author}{\bibfnamefont{L.~J.} \bibnamefont{Li}},
  \bibinfo{author}{\bibfnamefont{K.~P.} \bibnamefont{Loh}}, \bibnamefont{and}
  \bibinfo{author}{\bibfnamefont{H.}~\bibnamefont{Zhang}},
  \bibinfo{journal}{Nature Chemistry} \textbf{\bibinfo{volume}{5}},
  \bibinfo{pages}{263} (\bibinfo{year}{2013}),
  \urlprefix\url{https://doi.org/10.1038/nchem.1589}.

\bibitem[{\citenamefont{Qian et~al.}(2014)\citenamefont{Qian, Liu, Fu, and
  Li}}]{Qian2014}
\bibinfo{author}{\bibfnamefont{X.}~\bibnamefont{Qian}},
  \bibinfo{author}{\bibfnamefont{J.}~\bibnamefont{Liu}},
  \bibinfo{author}{\bibfnamefont{L.}~\bibnamefont{Fu}}, \bibnamefont{and}
  \bibinfo{author}{\bibfnamefont{J.}~\bibnamefont{Li}},
  \bibinfo{journal}{Science} \textbf{\bibinfo{volume}{346}},
  \bibinfo{pages}{1344} (\bibinfo{year}{2014}), ISSN \bibinfo{issn}{0036-8075},
  \urlprefix\url{https://science.sciencemag.org/content/346/6215/1344}.

\bibitem[{\citenamefont{Xiao et~al.}(2012)\citenamefont{Xiao, Liu, Feng, Xu,
  and Yao}}]{Xiao2012}
\bibinfo{author}{\bibfnamefont{D.}~\bibnamefont{Xiao}},
  \bibinfo{author}{\bibfnamefont{G.-B.} \bibnamefont{Liu}},
  \bibinfo{author}{\bibfnamefont{W.}~\bibnamefont{Feng}},
  \bibinfo{author}{\bibfnamefont{X.}~\bibnamefont{Xu}}, \bibnamefont{and}
  \bibinfo{author}{\bibfnamefont{W.}~\bibnamefont{Yao}},
  \bibinfo{journal}{Phys. Rev. Lett.} \textbf{\bibinfo{volume}{108}},
  \bibinfo{pages}{196802} (\bibinfo{year}{2012}),
  \urlprefix\url{https://link.aps.org/doi/10.1103/PhysRevLett.108.196802}.

\bibitem[{\citenamefont{Mak et~al.}(2012)\citenamefont{Mak, He, Shan, and
  Heinz}}]{Mak2012}
\bibinfo{author}{\bibfnamefont{K.~F.} \bibnamefont{Mak}},
  \bibinfo{author}{\bibfnamefont{K.}~\bibnamefont{He}},
  \bibinfo{author}{\bibfnamefont{J.}~\bibnamefont{Shan}}, \bibnamefont{and}
  \bibinfo{author}{\bibfnamefont{T.~F.} \bibnamefont{Heinz}},
  \bibinfo{journal}{Nature Nanotechnology} \textbf{\bibinfo{volume}{7}},
  \bibinfo{pages}{494} (\bibinfo{year}{2012}), ISSN \bibinfo{issn}{17483395},
  \eprint{1205.1822}, \urlprefix\url{https://doi.org/10.1038/nnano.2012.96}.

\bibitem[{\citenamefont{Ravindra et~al.}(2019)\citenamefont{Ravindra, Tang, and
  Rassay}}]{Ravindra2019}
\bibinfo{author}{\bibfnamefont{N.~M.} \bibnamefont{Ravindra}},
  \bibinfo{author}{\bibfnamefont{W.}~\bibnamefont{Tang}}, \bibnamefont{and}
  \bibinfo{author}{\bibfnamefont{S.}~\bibnamefont{Rassay}},
  \emph{\bibinfo{title}{Transition Metal Dichalcogenides Properties and
  Applications}} (\bibinfo{publisher}{Springer International Publishing},
  \bibinfo{address}{Cham}, \bibinfo{year}{2019}), pp.
  \bibinfo{pages}{333--396}, ISBN \bibinfo{isbn}{978-3-030-02171-9},
  \urlprefix\url{https://doi.org/10.1007/978-3-030-02171-9_6}.

\bibitem[{\citenamefont{Wang et~al.}(2012)\citenamefont{Wang, Kalantar-Zadeh,
  Kis, Coleman, and Strano}}]{Wang2012}
\bibinfo{author}{\bibfnamefont{Q.~H.} \bibnamefont{Wang}},
  \bibinfo{author}{\bibfnamefont{K.}~\bibnamefont{Kalantar-Zadeh}},
  \bibinfo{author}{\bibfnamefont{A.}~\bibnamefont{Kis}},
  \bibinfo{author}{\bibfnamefont{J.~N.} \bibnamefont{Coleman}},
  \bibnamefont{and} \bibinfo{author}{\bibfnamefont{M.~S.}
  \bibnamefont{Strano}}, \bibinfo{journal}{Nature Nanotechnology}
  \textbf{\bibinfo{volume}{7}}, \bibinfo{pages}{699} (\bibinfo{year}{2012}),
  ISSN \bibinfo{issn}{17483395},
  \urlprefix\url{https://doi.org/10.1038/nnano.2012.193}.

\bibitem[{\citenamefont{Lin et~al.}(2014)\citenamefont{Lin, Xu, Wang, Li,
  Yamamoto, Aparecido-Ferreira, Li, Sun, Nakaharai, Jian et~al.}}]{Lin2014}
\bibinfo{author}{\bibfnamefont{Y.~F.} \bibnamefont{Lin}},
  \bibinfo{author}{\bibfnamefont{Y.}~\bibnamefont{Xu}},
  \bibinfo{author}{\bibfnamefont{S.~T.} \bibnamefont{Wang}},
  \bibinfo{author}{\bibfnamefont{S.~L.} \bibnamefont{Li}},
  \bibinfo{author}{\bibfnamefont{M.}~\bibnamefont{Yamamoto}},
  \bibinfo{author}{\bibfnamefont{A.}~\bibnamefont{Aparecido-Ferreira}},
  \bibinfo{author}{\bibfnamefont{W.}~\bibnamefont{Li}},
  \bibinfo{author}{\bibfnamefont{H.}~\bibnamefont{Sun}},
  \bibinfo{author}{\bibfnamefont{S.}~\bibnamefont{Nakaharai}},
  \bibinfo{author}{\bibfnamefont{W.~B.} \bibnamefont{Jian}},
  \bibnamefont{et~al.}, \bibinfo{journal}{Advanced Materials}
  \textbf{\bibinfo{volume}{26}}, \bibinfo{pages}{3263} (\bibinfo{year}{2014}),
  \urlprefix\url{https://www.onlinelibrary.wiley.com/doi/abs/10.1002/adma.201305845}.

\bibitem[{\citenamefont{Hayat et~al.}(2016)\citenamefont{Hayat, Kohary, and
  Wright}}]{Hayat2017}
\bibinfo{author}{\bibfnamefont{H.}~\bibnamefont{Hayat}},
  \bibinfo{author}{\bibfnamefont{K.}~\bibnamefont{Kohary}}, \bibnamefont{and}
  \bibinfo{author}{\bibfnamefont{C.~D.} \bibnamefont{Wright}},
  \bibinfo{journal}{Nanotechnology} \textbf{\bibinfo{volume}{28}},
  \bibinfo{pages}{035202} (\bibinfo{year}{2016}),
  \urlprefix\url{https://doi.org/10.1088/1361-6528/28/3/035202}.

\bibitem[{\citenamefont{Duerloo et~al.}(2014)\citenamefont{Duerloo, Li, and
  Reed}}]{Duerloo2014}
\bibinfo{author}{\bibfnamefont{K.-A.~N.} \bibnamefont{Duerloo}},
  \bibinfo{author}{\bibfnamefont{Y.}~\bibnamefont{Li}}, \bibnamefont{and}
  \bibinfo{author}{\bibfnamefont{E.~J.} \bibnamefont{Reed}},
  \bibinfo{journal}{Nature Communications} \textbf{\bibinfo{volume}{5}},
  \bibinfo{pages}{4214} (\bibinfo{year}{2014}),
  \urlprefix\url{https://doi.org/10.1038/ncomms5214}.

\bibitem[{\citenamefont{Li et~al.}(2016)\citenamefont{Li, Duerloo, Wauson, and
  Reed}}]{Li2016}
\bibinfo{author}{\bibfnamefont{Y.}~\bibnamefont{Li}},
  \bibinfo{author}{\bibfnamefont{K.~A.~N.} \bibnamefont{Duerloo}},
  \bibinfo{author}{\bibfnamefont{K.}~\bibnamefont{Wauson}}, \bibnamefont{and}
  \bibinfo{author}{\bibfnamefont{E.~J.} \bibnamefont{Reed}},
  \bibinfo{journal}{Nature Communications} \textbf{\bibinfo{volume}{7}},
  \bibinfo{pages}{10671} (\bibinfo{year}{2016}).

\bibitem[{\citenamefont{Yang et~al.}(2017)\citenamefont{Yang, Kim, Chhowalla,
  and Lee}}]{Yang2017}
\bibinfo{author}{\bibfnamefont{H.}~\bibnamefont{Yang}},
  \bibinfo{author}{\bibfnamefont{S.~W.} \bibnamefont{Kim}},
  \bibinfo{author}{\bibfnamefont{M.}~\bibnamefont{Chhowalla}},
  \bibnamefont{and} \bibinfo{author}{\bibfnamefont{Y.~H.} \bibnamefont{Lee}},
  \bibinfo{journal}{Nature Physics} \textbf{\bibinfo{volume}{13}},
  \bibinfo{pages}{931} (\bibinfo{year}{2017}),
  \urlprefix\url{https://doi.org/10.1038/nphys4188}.

\bibitem[{\citenamefont{Rhodes et~al.}(2017)\citenamefont{Rhodes, Chenet,
  Janicek, Nyby, Lin, Jin, Edelberg, Mannebach, Finney, Antony
  et~al.}}]{Rhodes2017}
\bibinfo{author}{\bibfnamefont{D.}~\bibnamefont{Rhodes}},
  \bibinfo{author}{\bibfnamefont{D.~A.} \bibnamefont{Chenet}},
  \bibinfo{author}{\bibfnamefont{B.~E.} \bibnamefont{Janicek}},
  \bibinfo{author}{\bibfnamefont{C.}~\bibnamefont{Nyby}},
  \bibinfo{author}{\bibfnamefont{Y.}~\bibnamefont{Lin}},
  \bibinfo{author}{\bibfnamefont{W.}~\bibnamefont{Jin}},
  \bibinfo{author}{\bibfnamefont{D.}~\bibnamefont{Edelberg}},
  \bibinfo{author}{\bibfnamefont{E.}~\bibnamefont{Mannebach}},
  \bibinfo{author}{\bibfnamefont{N.}~\bibnamefont{Finney}},
  \bibinfo{author}{\bibfnamefont{A.}~\bibnamefont{Antony}},
  \bibnamefont{et~al.}, \bibinfo{journal}{Nano Letters}
  \textbf{\bibinfo{volume}{17}}, \bibinfo{pages}{1616} (\bibinfo{year}{2017}),
  \urlprefix\url{https://doi.org/10.1021/acs.nanolett.6b04814}.

\bibitem[{\citenamefont{Wang et~al.}(2017)\citenamefont{Wang, Xiao, Zhu, Li,
  Alsaid, Fong, Zhou, Wang, Shi, Wang et~al.}}]{Wang2017}
\bibinfo{author}{\bibfnamefont{Y.}~\bibnamefont{Wang}},
  \bibinfo{author}{\bibfnamefont{J.}~\bibnamefont{Xiao}},
  \bibinfo{author}{\bibfnamefont{H.}~\bibnamefont{Zhu}},
  \bibinfo{author}{\bibfnamefont{Y.}~\bibnamefont{Li}},
  \bibinfo{author}{\bibfnamefont{Y.}~\bibnamefont{Alsaid}},
  \bibinfo{author}{\bibfnamefont{K.~Y.} \bibnamefont{Fong}},
  \bibinfo{author}{\bibfnamefont{Y.}~\bibnamefont{Zhou}},
  \bibinfo{author}{\bibfnamefont{S.}~\bibnamefont{Wang}},
  \bibinfo{author}{\bibfnamefont{W.}~\bibnamefont{Shi}},
  \bibinfo{author}{\bibfnamefont{Y.}~\bibnamefont{Wang}}, \bibnamefont{et~al.},
  \bibinfo{journal}{Nature} \textbf{\bibinfo{volume}{550}},
  \bibinfo{pages}{487} (\bibinfo{year}{2017}).

\bibitem[{\citenamefont{Palacios-Berraquero
  et~al.}(2017)\citenamefont{Palacios-Berraquero, Kara, Montblanch, Barbone,
  Latawiec, Yoon, Ott, Loncar, Ferrari, and
  Atat\"{u}re}}]{Loncar_TMD_Nature_2017}
\bibinfo{author}{\bibfnamefont{C.}~\bibnamefont{Palacios-Berraquero}},
  \bibinfo{author}{\bibfnamefont{D.~M.} \bibnamefont{Kara}},
  \bibinfo{author}{\bibfnamefont{A.~R.-P.} \bibnamefont{Montblanch}},
  \bibinfo{author}{\bibfnamefont{M.}~\bibnamefont{Barbone}},
  \bibinfo{author}{\bibfnamefont{P.}~\bibnamefont{Latawiec}},
  \bibinfo{author}{\bibfnamefont{D.}~\bibnamefont{Yoon}},
  \bibinfo{author}{\bibfnamefont{A.~K.} \bibnamefont{Ott}},
  \bibinfo{author}{\bibfnamefont{M.}~\bibnamefont{Loncar}},
  \bibinfo{author}{\bibfnamefont{A.~C.} \bibnamefont{Ferrari}},
  \bibnamefont{and}
  \bibinfo{author}{\bibfnamefont{M.}~\bibnamefont{Atat\"{u}re}},
  \bibinfo{journal}{Nat. Commun.} \textbf{\bibinfo{volume}{8}},
  \bibinfo{pages}{15093} (\bibinfo{year}{2017}),
  \eprint{https://doi.org/10.1038/ncomms15093}.

\bibitem[{\citenamefont{Manchanda et~al.}(2020)\citenamefont{Manchanda, Kumar,
  and Dev}}]{Manchanda2020}
\bibinfo{author}{\bibfnamefont{P.}~\bibnamefont{Manchanda}},
  \bibinfo{author}{\bibfnamefont{P.}~\bibnamefont{Kumar}}, \bibnamefont{and}
  \bibinfo{author}{\bibfnamefont{P.}~\bibnamefont{Dev}},
  \bibinfo{journal}{Phys. Rev. B} \textbf{\bibinfo{volume}{101}},
  \bibinfo{pages}{144104} (\bibinfo{year}{2020}),
  \urlprefix\url{https://link.aps.org/doi/10.1103/PhysRevB.101.144104}.

\bibitem[{\citenamefont{Manchanda et~al.}(2021)\citenamefont{Manchanda, Kumar,
  and Dev}}]{Manchanda_PtSe2_2021}
\bibinfo{author}{\bibfnamefont{P.}~\bibnamefont{Manchanda}},
  \bibinfo{author}{\bibfnamefont{P.}~\bibnamefont{Kumar}}, \bibnamefont{and}
  \bibinfo{author}{\bibfnamefont{P.}~\bibnamefont{Dev}},
  \bibinfo{journal}{Phys. Rev. B} \textbf{\bibinfo{volume}{103}},
  \bibinfo{pages}{144403} (\bibinfo{year}{2021}),
  \urlprefix\url{https://link.aps.org/doi/10.1103/PhysRevB.103.144403}.

\bibitem[{\citenamefont{Lu et~al.}(2017)\citenamefont{Lu, Zhu, Xiao, Chuu, Han,
  Chiu, Cheng, Yang, Wei, Yang et~al.}}]{Lu2017}
\bibinfo{author}{\bibfnamefont{A.~Y.} \bibnamefont{Lu}},
  \bibinfo{author}{\bibfnamefont{H.}~\bibnamefont{Zhu}},
  \bibinfo{author}{\bibfnamefont{J.}~\bibnamefont{Xiao}},
  \bibinfo{author}{\bibfnamefont{C.~P.} \bibnamefont{Chuu}},
  \bibinfo{author}{\bibfnamefont{Y.}~\bibnamefont{Han}},
  \bibinfo{author}{\bibfnamefont{M.~H.} \bibnamefont{Chiu}},
  \bibinfo{author}{\bibfnamefont{C.~C.} \bibnamefont{Cheng}},
  \bibinfo{author}{\bibfnamefont{C.~W.} \bibnamefont{Yang}},
  \bibinfo{author}{\bibfnamefont{K.~H.} \bibnamefont{Wei}},
  \bibinfo{author}{\bibfnamefont{Y.}~\bibnamefont{Yang}}, \bibnamefont{et~al.},
  \bibinfo{journal}{Nature Nanotechnology} \textbf{\bibinfo{volume}{12}},
  \bibinfo{pages}{744} (\bibinfo{year}{2017}), ISSN \bibinfo{issn}{17483395},
  \urlprefix\url{https://doi.org/10.1038/nnano.2017.100}.

\bibitem[{\citenamefont{Zhang et~al.}(2017)\citenamefont{Zhang, Jia, Kholmanov,
  Dong, Er, Chen, Guo, Jin, Shenoy, Shi et~al.}}]{MoSSe_Zhang_2017}
\bibinfo{author}{\bibfnamefont{J.}~\bibnamefont{Zhang}},
  \bibinfo{author}{\bibfnamefont{S.}~\bibnamefont{Jia}},
  \bibinfo{author}{\bibfnamefont{I.}~\bibnamefont{Kholmanov}},
  \bibinfo{author}{\bibfnamefont{L.}~\bibnamefont{Dong}},
  \bibinfo{author}{\bibfnamefont{D.}~\bibnamefont{Er}},
  \bibinfo{author}{\bibfnamefont{W.}~\bibnamefont{Chen}},
  \bibinfo{author}{\bibfnamefont{H.}~\bibnamefont{Guo}},
  \bibinfo{author}{\bibfnamefont{Z.}~\bibnamefont{Jin}},
  \bibinfo{author}{\bibfnamefont{V.~B.} \bibnamefont{Shenoy}},
  \bibinfo{author}{\bibfnamefont{L.}~\bibnamefont{Shi}}, \bibnamefont{et~al.},
  \bibinfo{journal}{ACS Nano} \textbf{\bibinfo{volume}{11}},
  \bibinfo{pages}{8192} (\bibinfo{year}{2017}), \bibinfo{note}{pMID: 28771310},
  \urlprefix\url{https://doi.org/10.1021/acsnano.7b03186}.

\bibitem[{\citenamefont{Zhang et~al.}(2019)\citenamefont{Zhang, Nie, Sanvito,
  and Du}}]{Zhang_VSSe_2019}
\bibinfo{author}{\bibfnamefont{C.}~\bibnamefont{Zhang}},
  \bibinfo{author}{\bibfnamefont{Y.}~\bibnamefont{Nie}},
  \bibinfo{author}{\bibfnamefont{S.}~\bibnamefont{Sanvito}}, \bibnamefont{and}
  \bibinfo{author}{\bibfnamefont{A.}~\bibnamefont{Du}}, \bibinfo{journal}{Nano
  Letters} \textbf{\bibinfo{volume}{19}}, \bibinfo{pages}{1366}
  (\bibinfo{year}{2019}),
  \urlprefix\url{https://doi.org/10.1021/acs.nanolett.8b05050}.

\bibitem[{\citenamefont{Pauling}(1929)}]{Pauling1929}
\bibinfo{author}{\bibfnamefont{L.}~\bibnamefont{Pauling}},
  \bibinfo{journal}{Journal of the American Chemical Society}
  \textbf{\bibinfo{volume}{51}}, \bibinfo{pages}{1010} (\bibinfo{year}{1929}),
  \urlprefix\url{https://doi.org/10.1021/ja01379a006}.

\bibitem[{sup()}]{supp}
\bibinfo{note}{See Supplemental Material at [URL will be inserted by the
  production group] for (i) a note on efficacy of combining high-throughput
  quantum mechanical computations with machine learning algorithms, (ii)
  additional RF regression and classification results for different feature
  sets, as well as KRR results, and (iii) tables with details from
  high-throughput DFT calculations and ML analysis, including different
  robustness tests for the ML models.}

\bibitem[{\citenamefont{Huan et~al.}(2016)\citenamefont{Huan,
  Mannodi-Kanakkithodi, Kim, Sharma, Pilania, and Ramprasad}}]{Huan2016}
\bibinfo{author}{\bibfnamefont{T.~D.} \bibnamefont{Huan}},
  \bibinfo{author}{\bibfnamefont{A.}~\bibnamefont{Mannodi-Kanakkithodi}},
  \bibinfo{author}{\bibfnamefont{C.}~\bibnamefont{Kim}},
  \bibinfo{author}{\bibfnamefont{V.}~\bibnamefont{Sharma}},
  \bibinfo{author}{\bibfnamefont{G.}~\bibnamefont{Pilania}}, \bibnamefont{and}
  \bibinfo{author}{\bibfnamefont{R.}~\bibnamefont{Ramprasad}},
  \bibinfo{journal}{Scientific Data} \textbf{\bibinfo{volume}{3}},
  \bibinfo{pages}{160012} (\bibinfo{year}{2016}), ISSN
  \bibinfo{issn}{2052-4463},
  \urlprefix\url{https://doi.org/10.1038/sdata.2016.12}.

\bibitem[{\citenamefont{Tawfik et~al.}(2019)\citenamefont{Tawfik, Isayev,
  Stampfl, Shapter, Winkler, and Ford}}]{Tawfik2019}
\bibinfo{author}{\bibfnamefont{S.~A.} \bibnamefont{Tawfik}},
  \bibinfo{author}{\bibfnamefont{O.}~\bibnamefont{Isayev}},
  \bibinfo{author}{\bibfnamefont{C.}~\bibnamefont{Stampfl}},
  \bibinfo{author}{\bibfnamefont{J.}~\bibnamefont{Shapter}},
  \bibinfo{author}{\bibfnamefont{D.~A.} \bibnamefont{Winkler}},
  \bibnamefont{and} \bibinfo{author}{\bibfnamefont{M.~J.} \bibnamefont{Ford}},
  \bibinfo{journal}{Advanced Theory and Simulations}
  \textbf{\bibinfo{volume}{2}}, \bibinfo{pages}{1800128}
  (\bibinfo{year}{2019}), ISSN \bibinfo{issn}{2513-0390},
  \urlprefix\url{http://doi.wiley.com/10.1002/adts.201800128}.

\bibitem[{\citenamefont{Sharma et~al.}(2020)\citenamefont{Sharma, Kumar, Dev,
  and Pilania}}]{Sharma2020}
\bibinfo{author}{\bibfnamefont{V.}~\bibnamefont{Sharma}},
  \bibinfo{author}{\bibfnamefont{P.}~\bibnamefont{Kumar}},
  \bibinfo{author}{\bibfnamefont{P.}~\bibnamefont{Dev}}, \bibnamefont{and}
  \bibinfo{author}{\bibfnamefont{G.}~\bibnamefont{Pilania}},
  \bibinfo{journal}{J. Appl. Phys} \textbf{\bibinfo{volume}{128}},
  \bibinfo{pages}{34902} (\bibinfo{year}{2020}),
  \urlprefix\url{https://doi.org/10.1063/5.0015538}.

\bibitem[{\citenamefont{Masubuchi et~al.}(2020)\citenamefont{Masubuchi,
  Watanabe, Seo, Okazaki, Sasagawa, Watanabe, Taniguchi, and
  Machida}}]{Masubuchi2020}
\bibinfo{author}{\bibfnamefont{S.}~\bibnamefont{Masubuchi}},
  \bibinfo{author}{\bibfnamefont{E.}~\bibnamefont{Watanabe}},
  \bibinfo{author}{\bibfnamefont{Y.}~\bibnamefont{Seo}},
  \bibinfo{author}{\bibfnamefont{S.}~\bibnamefont{Okazaki}},
  \bibinfo{author}{\bibfnamefont{T.}~\bibnamefont{Sasagawa}},
  \bibinfo{author}{\bibfnamefont{K.}~\bibnamefont{Watanabe}},
  \bibinfo{author}{\bibfnamefont{T.}~\bibnamefont{Taniguchi}},
  \bibnamefont{and} \bibinfo{author}{\bibfnamefont{T.}~\bibnamefont{Machida}},
  \bibinfo{journal}{npj 2D Materials and Applications}
  \textbf{\bibinfo{volume}{4}}, \bibinfo{pages}{3} (\bibinfo{year}{2020}), ISSN
  \bibinfo{issn}{23977132},
  \urlprefix\url{https://doi.org/10.1038/s41699-020-0137-z}.

\bibitem[{\citenamefont{Schleder et~al.}(2020)\citenamefont{Schleder, Acosta,
  and Fazzio}}]{Schleder2020}
\bibinfo{author}{\bibfnamefont{G.~R.} \bibnamefont{Schleder}},
  \bibinfo{author}{\bibfnamefont{C.~M.} \bibnamefont{Acosta}},
  \bibnamefont{and} \bibinfo{author}{\bibfnamefont{A.}~\bibnamefont{Fazzio}},
  \bibinfo{journal}{ACS Applied Materials and Interfaces}
  \textbf{\bibinfo{volume}{12}}, \bibinfo{pages}{20149} (\bibinfo{year}{2020}),
  \urlprefix\url{https://doi.org/10.1021/acsami.9b14530}.

\bibitem[{\citenamefont{Choudhary et~al.}(2020)\citenamefont{Choudhary,
  Garrity, Sharma, Biacchi, Hight~Walker, and Tavazza}}]{Choudhary2020}
\bibinfo{author}{\bibfnamefont{K.}~\bibnamefont{Choudhary}},
  \bibinfo{author}{\bibfnamefont{K.~F.} \bibnamefont{Garrity}},
  \bibinfo{author}{\bibfnamefont{V.}~\bibnamefont{Sharma}},
  \bibinfo{author}{\bibfnamefont{A.~J.} \bibnamefont{Biacchi}},
  \bibinfo{author}{\bibfnamefont{A.~R.} \bibnamefont{Hight~Walker}},
  \bibnamefont{and} \bibinfo{author}{\bibfnamefont{F.}~\bibnamefont{Tavazza}},
  \bibinfo{journal}{npj Computational Materials} \textbf{\bibinfo{volume}{6}},
  \bibinfo{pages}{64} (\bibinfo{year}{2020}),
  \urlprefix\url{https://doi.org/10.1038/s41524-020-0337-2}.

\bibitem[{\citenamefont{{Choudhary} et~al.}(2020)\citenamefont{{Choudhary},
  {Garrity}, {Reid}, {DeCost}, {Biacchi}, {Hight Walker}, {Trautt},
  {Hattrick-Simpers}, {Gilad Kusne}, {Centrone} et~al.}}]{Jarvis2020}
\bibinfo{author}{\bibfnamefont{K.}~\bibnamefont{{Choudhary}}},
  \bibinfo{author}{\bibfnamefont{K.~F.} \bibnamefont{{Garrity}}},
  \bibinfo{author}{\bibfnamefont{A.~C.~E.} \bibnamefont{{Reid}}},
  \bibinfo{author}{\bibfnamefont{B.}~\bibnamefont{{DeCost}}},
  \bibinfo{author}{\bibfnamefont{A.~J.} \bibnamefont{{Biacchi}}},
  \bibinfo{author}{\bibfnamefont{A.~R.} \bibnamefont{{Hight Walker}}},
  \bibinfo{author}{\bibfnamefont{Z.}~\bibnamefont{{Trautt}}},
  \bibinfo{author}{\bibfnamefont{J.}~\bibnamefont{{Hattrick-Simpers}}},
  \bibinfo{author}{\bibfnamefont{A.}~\bibnamefont{{Gilad Kusne}}},
  \bibinfo{author}{\bibfnamefont{A.}~\bibnamefont{{Centrone}}},
  \bibnamefont{et~al.}, \bibinfo{journal}{npj Computational Materials}
  \textbf{\bibinfo{volume}{6}}, \bibinfo{pages}{173} (\bibinfo{year}{2020}),
  \urlprefix\url{https://doi.org/10.1038/s41524-020-00440-1}.

\bibitem[{\citenamefont{Kresse and Hafner}(1994)}]{Kresse1994}
\bibinfo{author}{\bibfnamefont{G.}~\bibnamefont{Kresse}} \bibnamefont{and}
  \bibinfo{author}{\bibfnamefont{J.}~\bibnamefont{Hafner}},
  \bibinfo{journal}{Phys. Rev. B} \textbf{\bibinfo{volume}{49}},
  \bibinfo{pages}{14251} (\bibinfo{year}{1994}),
  \urlprefix\url{https://link.aps.org/doi/10.1103/PhysRevB.49.14251}.

\bibitem[{\citenamefont{Kresse and Furthm{\"u}ller}(1996)}]{Kresse1996}
\bibinfo{author}{\bibfnamefont{G.}~\bibnamefont{Kresse}} \bibnamefont{and}
  \bibinfo{author}{\bibfnamefont{J.}~\bibnamefont{Furthm{\"u}ller}},
  \bibinfo{journal}{Computational Materials Science}
  \textbf{\bibinfo{volume}{6}}, \bibinfo{pages}{15} (\bibinfo{year}{1996}),
  ISSN \bibinfo{issn}{0927-0256},
  \urlprefix\url{https://www.sciencedirect.com/science/article/pii/0927025696000080}.

\bibitem[{\citenamefont{Perdew et~al.}(1996)\citenamefont{Perdew, Burke, and
  Ernzerhof}}]{Perdew1996}
\bibinfo{author}{\bibfnamefont{J.~P.} \bibnamefont{Perdew}},
  \bibinfo{author}{\bibfnamefont{K.}~\bibnamefont{Burke}}, \bibnamefont{and}
  \bibinfo{author}{\bibfnamefont{M.}~\bibnamefont{Ernzerhof}},
  \bibinfo{journal}{Phys. Rev. Lett.} \textbf{\bibinfo{volume}{77}},
  \bibinfo{pages}{3865} (\bibinfo{year}{1996}),
  \urlprefix\url{https://link.aps.org/doi/10.1103/PhysRevLett.77.3865}.

\bibitem[{\citenamefont{Dev et~al.}(2008)\citenamefont{Dev, Xue, and
  Zhang}}]{Dev_PRL_2008}
\bibinfo{author}{\bibfnamefont{P.}~\bibnamefont{Dev}},
  \bibinfo{author}{\bibfnamefont{Y.}~\bibnamefont{Xue}}, \bibnamefont{and}
  \bibinfo{author}{\bibfnamefont{P.}~\bibnamefont{Zhang}},
  \bibinfo{journal}{Phys. Rev. Lett.} \textbf{\bibinfo{volume}{100}},
  \bibinfo{pages}{117204} (\bibinfo{year}{2008}),
  \urlprefix\url{https://link.aps.org/doi/10.1103/PhysRevLett.100.117204}.

\bibitem[{\citenamefont{Dev and Zhang}(2010)}]{Dev_PRB_2010}
\bibinfo{author}{\bibfnamefont{P.}~\bibnamefont{Dev}} \bibnamefont{and}
  \bibinfo{author}{\bibfnamefont{P.}~\bibnamefont{Zhang}},
  \bibinfo{journal}{Phys. Rev. B} \textbf{\bibinfo{volume}{81}},
  \bibinfo{pages}{085207} (\bibinfo{year}{2010}),
  \urlprefix\url{https://link.aps.org/doi/10.1103/PhysRevB.81.085207}.

\bibitem[{\citenamefont{Monkhorst and Pack}(1976)}]{Monkhorst1976}
\bibinfo{author}{\bibfnamefont{H.~J.} \bibnamefont{Monkhorst}}
  \bibnamefont{and} \bibinfo{author}{\bibfnamefont{J.~D.} \bibnamefont{Pack}},
  \bibinfo{journal}{Phys. Rev. B} \textbf{\bibinfo{volume}{13}},
  \bibinfo{pages}{5188} (\bibinfo{year}{1976}),
  \urlprefix\url{https://link.aps.org/doi/10.1103/PhysRevB.13.5188}.

\bibitem[{\citenamefont{Shannon}(1976)}]{Shannon1976}
\bibinfo{author}{\bibfnamefont{R.~D.} \bibnamefont{Shannon}},
  \bibinfo{journal}{Acta Crystallographica Section A}
  \textbf{\bibinfo{volume}{32}}, \bibinfo{pages}{751} (\bibinfo{year}{1976}),
  \urlprefix\url{https://www.onlinelibrary.wiley.com/doi/abs/10.1107/S0567739476001551}.

\bibitem[{men(2014--)}]{mendeleev2014}
\emph{\bibinfo{title}{{mendeleev} -- a python resource for properties of
  chemical elements, ions and isotopes, ver. 0.6.0}},
  \bibinfo{howpublished}{\url{https://github.com/lmmentel/mendeleev}}
  (\bibinfo{year}{2014--}).

\bibitem[{\citenamefont{Breiman}(2001)}]{Breiman2001}
\bibinfo{author}{\bibfnamefont{L.}~\bibnamefont{Breiman}},
  \bibinfo{journal}{Machine Learning} \textbf{\bibinfo{volume}{45}},
  \bibinfo{pages}{5} (\bibinfo{year}{2001}),
  \urlprefix\url{https://doi.org/10.1023/A:1010933404324}.

\bibitem[{\citenamefont{Zhang and Ma}(2012)}]{ensemble}
\bibinfo{author}{\bibfnamefont{C.}~\bibnamefont{Zhang}} \bibnamefont{and}
  \bibinfo{author}{\bibfnamefont{Y.}~\bibnamefont{Ma}},
  \emph{\bibinfo{title}{{Ensemble Machine Learning}}}
  (\bibinfo{publisher}{Springer, Boston, MA}, \bibinfo{year}{2012}), ISBN
  \bibinfo{isbn}{978-1-4419-9326-7}.

\bibitem[{\citenamefont{Ataca et~al.}(2012)\citenamefont{Ataca, ?ahin, and
  Ciraci}}]{Ataca2012}
\bibinfo{author}{\bibfnamefont{C.}~\bibnamefont{Ataca}},
  \bibinfo{author}{\bibfnamefont{H.}~\bibnamefont{?ahin}}, \bibnamefont{and}
  \bibinfo{author}{\bibfnamefont{S.}~\bibnamefont{Ciraci}},
  \bibinfo{journal}{The Journal of Physical Chemistry C}
  \textbf{\bibinfo{volume}{116}}, \bibinfo{pages}{8983} (\bibinfo{year}{2012}),
  \urlprefix\url{https://doi.org/10.1021/jp212558p}.

\bibitem[{\citenamefont{Rasmussen and Thygesen}(2015)}]{Rasmussen2015}
\bibinfo{author}{\bibfnamefont{F.~A.} \bibnamefont{Rasmussen}}
  \bibnamefont{and} \bibinfo{author}{\bibfnamefont{K.~S.}
  \bibnamefont{Thygesen}}, \bibinfo{journal}{The Journal of Physical Chemistry
  C} \textbf{\bibinfo{volume}{119}}, \bibinfo{pages}{13169}
  (\bibinfo{year}{2015}),
  \urlprefix\url{https://doi.org/10.1021/acs.jpcc.5b02950}.

\bibitem[{\citenamefont{Gao et~al.}(2013)\citenamefont{Gao, Xue, Mao, Wang, Xu,
  and Xue}}]{Gao2013}
\bibinfo{author}{\bibfnamefont{D.}~\bibnamefont{Gao}},
  \bibinfo{author}{\bibfnamefont{Q.}~\bibnamefont{Xue}},
  \bibinfo{author}{\bibfnamefont{X.}~\bibnamefont{Mao}},
  \bibinfo{author}{\bibfnamefont{W.}~\bibnamefont{Wang}},
  \bibinfo{author}{\bibfnamefont{Q.}~\bibnamefont{Xu}}, \bibnamefont{and}
  \bibinfo{author}{\bibfnamefont{D.}~\bibnamefont{Xue}}, \bibinfo{journal}{J.
  Mater. Chem. C} \textbf{\bibinfo{volume}{1}}, \bibinfo{pages}{5909}
  (\bibinfo{year}{2013}), \urlprefix\url{http://dx.doi.org/10.1039/C3TC31233J}.

\bibitem[{\citenamefont{Zhou et~al.}(2018)\citenamefont{Zhou, Lin, Huang, Zhou,
  Chen, Xia, Wang, Xie, Yu, Lei et~al.}}]{Zhou2018}
\bibinfo{author}{\bibfnamefont{J.}~\bibnamefont{Zhou}},
  \bibinfo{author}{\bibfnamefont{J.}~\bibnamefont{Lin}},
  \bibinfo{author}{\bibfnamefont{X.}~\bibnamefont{Huang}},
  \bibinfo{author}{\bibfnamefont{Y.}~\bibnamefont{Zhou}},
  \bibinfo{author}{\bibfnamefont{Y.}~\bibnamefont{Chen}},
  \bibinfo{author}{\bibfnamefont{J.}~\bibnamefont{Xia}},
  \bibinfo{author}{\bibfnamefont{H.}~\bibnamefont{Wang}},
  \bibinfo{author}{\bibfnamefont{Y.}~\bibnamefont{Xie}},
  \bibinfo{author}{\bibfnamefont{H.}~\bibnamefont{Yu}},
  \bibinfo{author}{\bibfnamefont{J.}~\bibnamefont{Lei}}, \bibnamefont{et~al.},
  \bibinfo{journal}{Nature} \textbf{\bibinfo{volume}{556}},
  \bibinfo{pages}{355} (\bibinfo{year}{2018}),
  \urlprefix\url{https://doi.org/10.1038/s41586-018-0008-3}.

\bibitem[{\citenamefont{Lu and Sinnott}(2020)}]{Lu2019}
\bibinfo{author}{\bibfnamefont{Y.}~\bibnamefont{Lu}} \bibnamefont{and}
  \bibinfo{author}{\bibfnamefont{S.~B.} \bibnamefont{Sinnott}},
  \bibinfo{journal}{ACS Applied Nano Materials} \textbf{\bibinfo{volume}{3}},
  \bibinfo{pages}{384} (\bibinfo{year}{2020}),
  \urlprefix\url{https://doi.org/10.1021/acsanm.9b02021}.

\bibitem[{\citenamefont{O?Hara et~al.}(2018)\citenamefont{O?Hara, Zhu,
  Trout, Ahmed, Luo, Lee, Brenner, Rajan, Gupta, McComb et~al.}}]{OHara2018}
\bibinfo{author}{\bibfnamefont{D.~J.} \bibnamefont{O?Hara}},
  \bibinfo{author}{\bibfnamefont{T.}~\bibnamefont{Zhu}},
  \bibinfo{author}{\bibfnamefont{A.~H.} \bibnamefont{Trout}},
  \bibinfo{author}{\bibfnamefont{A.~S.} \bibnamefont{Ahmed}},
  \bibinfo{author}{\bibfnamefont{Y.~K.} \bibnamefont{Luo}},
  \bibinfo{author}{\bibfnamefont{C.~H.} \bibnamefont{Lee}},
  \bibinfo{author}{\bibfnamefont{M.~R.} \bibnamefont{Brenner}},
  \bibinfo{author}{\bibfnamefont{S.}~\bibnamefont{Rajan}},
  \bibinfo{author}{\bibfnamefont{J.~A.} \bibnamefont{Gupta}},
  \bibinfo{author}{\bibfnamefont{D.~W.} \bibnamefont{McComb}},
  \bibnamefont{et~al.}, \bibinfo{journal}{Nano Letters}
  \textbf{\bibinfo{volume}{18}}, \bibinfo{pages}{3125} (\bibinfo{year}{2018}),
  \bibinfo{note}{pMID: 29608316},
  \urlprefix\url{https://doi.org/10.1021/acs.nanolett.8b00683}.

\bibitem[{\citenamefont{Kanazawa et~al.}(2016)\citenamefont{Kanazawa, Amemiya,
  Ishikawa, Upadhyaya, Tsuruta, Tanaka, and Miyamoto}}]{Kanazawa2016}
\bibinfo{author}{\bibfnamefont{T.}~\bibnamefont{Kanazawa}},
  \bibinfo{author}{\bibfnamefont{T.}~\bibnamefont{Amemiya}},
  \bibinfo{author}{\bibfnamefont{A.}~\bibnamefont{Ishikawa}},
  \bibinfo{author}{\bibfnamefont{V.}~\bibnamefont{Upadhyaya}},
  \bibinfo{author}{\bibfnamefont{K.}~\bibnamefont{Tsuruta}},
  \bibinfo{author}{\bibfnamefont{T.}~\bibnamefont{Tanaka}}, \bibnamefont{and}
  \bibinfo{author}{\bibfnamefont{Y.}~\bibnamefont{Miyamoto}},
  \bibinfo{journal}{Scientific Reports} \textbf{\bibinfo{volume}{6}},
  \bibinfo{pages}{22277} (\bibinfo{year}{2016}),
  \urlprefix\url{https://doi.org/10.1038/srep22277}.

\bibitem[{\citenamefont{Mounet et~al.}(2018)\citenamefont{Mounet, Gibertini,
  Schwaller, Campi, Merkys, Marrazzo, Sohier, Castelli, Cepellotti, Pizzi
  et~al.}}]{Mounet_Marzari2018}
\bibinfo{author}{\bibfnamefont{N.}~\bibnamefont{Mounet}},
  \bibinfo{author}{\bibfnamefont{M.}~\bibnamefont{Gibertini}},
  \bibinfo{author}{\bibfnamefont{P.}~\bibnamefont{Schwaller}},
  \bibinfo{author}{\bibfnamefont{D.}~\bibnamefont{Campi}},
  \bibinfo{author}{\bibfnamefont{A.}~\bibnamefont{Merkys}},
  \bibinfo{author}{\bibfnamefont{A.}~\bibnamefont{Marrazzo}},
  \bibinfo{author}{\bibfnamefont{T.}~\bibnamefont{Sohier}},
  \bibinfo{author}{\bibfnamefont{I.~E.} \bibnamefont{Castelli}},
  \bibinfo{author}{\bibfnamefont{A.}~\bibnamefont{Cepellotti}},
  \bibinfo{author}{\bibfnamefont{G.}~\bibnamefont{Pizzi}},
  \bibnamefont{et~al.}, \bibinfo{journal}{Nature Nanotechnology}
  \textbf{\bibinfo{volume}{13}}, \bibinfo{pages}{246} (\bibinfo{year}{2018}),
  \urlprefix\url{https://doi.org/10.1038/s41565-017-0035-5}.

\bibitem[{\citenamefont{C et~al.}(2015)\citenamefont{C, Zhang, Hong, Wallace,
  and Cho}}]{Santosh2015}
\bibinfo{author}{\bibfnamefont{S.~K.} \bibnamefont{C}},
  \bibinfo{author}{\bibfnamefont{C.}~\bibnamefont{Zhang}},
  \bibinfo{author}{\bibfnamefont{S.}~\bibnamefont{Hong}},
  \bibinfo{author}{\bibfnamefont{R.~M.} \bibnamefont{Wallace}},
  \bibnamefont{and} \bibinfo{author}{\bibfnamefont{K.}~\bibnamefont{Cho}},
  \bibinfo{journal}{2D Materials} \textbf{\bibinfo{volume}{2}},
  \bibinfo{pages}{035019} (\bibinfo{year}{2015}),
  \urlprefix\url{https://doi.org/10.1088%2F2053-1583%2F2%2F3%2F035019}.

\bibitem[{\citenamefont{Zhong et~al.}(2015)\citenamefont{Zhong, Gao, Shi, and
  Yang}}]{Zhong2015}
\bibinfo{author}{\bibfnamefont{H.-X.} \bibnamefont{Zhong}},
  \bibinfo{author}{\bibfnamefont{S.}~\bibnamefont{Gao}},
  \bibinfo{author}{\bibfnamefont{J.-J.} \bibnamefont{Shi}}, \bibnamefont{and}
  \bibinfo{author}{\bibfnamefont{L.}~\bibnamefont{Yang}},
  \bibinfo{journal}{Phys. Rev. B} \textbf{\bibinfo{volume}{92}},
  \bibinfo{pages}{115438} (\bibinfo{year}{2015}),
  \urlprefix\url{https://link.aps.org/doi/10.1103/PhysRevB.92.115438}.

\bibitem[{\citenamefont{Huisman et~al.}(1971)\citenamefont{Huisman, {de Jonge},
  Haas, and Jellinek}}]{Huisman1971}
\bibinfo{author}{\bibfnamefont{R.}~\bibnamefont{Huisman}},
  \bibinfo{author}{\bibfnamefont{R.}~\bibnamefont{{de Jonge}}},
  \bibinfo{author}{\bibfnamefont{C.}~\bibnamefont{Haas}}, \bibnamefont{and}
  \bibinfo{author}{\bibfnamefont{F.}~\bibnamefont{Jellinek}},
  \bibinfo{journal}{Journal of Solid State Chemistry}
  \textbf{\bibinfo{volume}{3}}, \bibinfo{pages}{56} (\bibinfo{year}{1971}),
  ISSN \bibinfo{issn}{0022-4596},
  \urlprefix\url{https://www.sciencedirect.com/science/article/pii/0022459671900077}.

\end{thebibliography}

\end{document}